\documentclass[aos]{imsart}

\RequirePackage{amsthm,amsmath,amsfonts,amssymb,amstext}
\RequirePackage[authoryear]{natbib}
\RequirePackage{bm}
\RequirePackage{booktabs} 
\RequirePackage{color}
\RequirePackage{enumerate} 
\RequirePackage[shortlabels]{enumitem}
\RequirePackage{epsfig,epsf,psfrag}
\RequirePackage{epstopdf}
\RequirePackage{float}
\RequirePackage{graphics}
\RequirePackage{graphicx}
\RequirePackage[colorlinks,linkcolor=black,citecolor={blue!50!black},urlcolor={blue!50!black}]{hyperref}
\RequirePackage{latexsym}
\RequirePackage{lscape}
\RequirePackage{mathabx} 
\RequirePackage{mathtools}
\RequirePackage{mathrsfs}
\RequirePackage{multirow}
\RequirePackage{natbib}
\RequirePackage{rotate}
\RequirePackage{xargs}[2008/03/08]
\RequirePackage{xcolor}
\RequirePackage{xspace}
\RequirePackage{xparse}

\startlocaldefs

\newtheorem{theorem}{Theorem}

\newtheorem{proposition}[theorem]{Proposition}
\newtheorem{corollary}[theorem]{Corollary}
{
	\theoremstyle{remark}
	\newtheorem{remark}{Remark}
	\newtheorem{example}{Example}
    \newtheorem{assumption}{Assumption}
    \newtheorem{assumptionalt}{Assumption}
}

\newenvironment{assumptionp}[1]{
  \renewcommand\theassumptionalt{#1}
  \assumptionalt
}{\endassumptionalt}

\newtheorem*{lemma*}{Lemma}

\definecolor{DarkBlue}{rgb}{0,.08,.45}
\definecolor{DarkRed}{rgb}{.7,0,.4}
\definecolor{DarkGreen}{rgb}{0.025,.5,.17}

\newcommand{\bea}{\begin{eqnarray*}}
\newcommand{\eea}{\end{eqnarray*}}
\newcommand{\be}{\begin{eqnarray}}
\newcommand{\ee}{\end{eqnarray}}
\newcommand{\beq}{\begin{equation}}
\newcommand{\eeq}{\end{equation}}

\newcommand{\bgt}{\begin{equation}\begin{gathered}}
\newcommand{\egt}{\end{gathered}\end{equation}}
\newcommand{\bal}{\begin{equation}\aligned}
\newcommand{\eal}{\endaligned\end{equation}}
\newcommand{\ed}{\end{document}}

\newcommand{\btab}{\begin{tabular}}
\newcommand{\etab}{\end{tabular}}

\newcommand{\bi}{\begin{itemize}}
\newcommand{\ei}{\end{itemize}}
\newcommand{\bfi}{\begin{figure}}
\newcommand{\efi}{\end{figure}}
\newcommand{\ben}{\begin{enumerate}}
\newcommand{\een}{\end{enumerate}}
\newcommand{\bay}{\begin{array}}
\newcommand{\eay}{\end{array}}

\newcommand{\nn}{\nonumber}

\newcommand{\bc}{\begin{center}}
\newcommand{\ec}{\end{center}}

\def\diag{{\rm diag}}

\def\diam{{\rm diam}}
\def\supp{\mathrm{supp}} 

\def\tps{^\mathrm{T}}
\def\indicator{\mathbb{I}}

\def\eps{\varepsilon}

\def\id{{\rm id}}

\def\diffop{\mathrm{d}}

\def\inv{^{-1}}
\def\cond{\mid} 

\def\wh{\widehat}
\def\wt{\widetilde}

\def\ra{\rightarrow}

\def \mbb {\mathbb}

\def \mc {\mathcal}
\def \msc{\mathscr}

\def\borel{\msc{B}}
\def\E{\mbb{E}} 
\def\prob{\mbb{P}} 
\def\real{\mbb{R}}

\def\Op{O_p}

\newcounter{const}
\NewDocumentCommand{\constant}{o}
 {
  \IfValueTF{#1}%
  {\ensuremath{C_{#1}}}%
  {\refstepcounter{const}%
  \ensuremath{C_{\theconst}}}%
 }

\newcounter{fun}
\NewDocumentCommand{\function}{oo
}{
  \IfValueTF{#2}%
  {\ensuremath{f_{\substack{#2,#1}}}}%
  {\refstepcounter{fun}%
  \ensuremath{f_{\substack{\thefun,#1}}}}%
 }

\newcounter{fcl}
\NewDocumentCommand{\fclass}{o}
 {
  \IfValueTF{#1}%
  {\ensuremath{\mathcal{F}_{#1}}}%
  {\refstepcounter{fcl}%
  \ensuremath{\mathcal{F}_{\thefcl}}}%
 }

\def\jt{\textrm{jt}\xspace}
\def\jtAD{\textrm{jt-AD}\xspace}
\def\jtF{\textrm{jt-F}\xspace}
\def\product{\textrm{prod}\xspace}
\def\productAD{\textrm{prod-AD}\xspace}
\def\productF{\textrm{prod-F}\xspace}
\def\dcov{\textrm{dCov}\xspace}
\def\mdcov{\textrm{mdCov}\xspace}
\def\dhsic{\textrm{dHSIC}\xspace}
\def\rvec{V}
\def\rvecbm{\bm{V}}
\newcommand{\gc}[2]{\mathcal{G}_{\mathrm{cop}}\{\mathbf{0},#1,#2\}}
\def\degC{$^{\circ}\mathrm{C}$}

\def\references{\bibliography{merged}}
\graphicspath{{figs/}}

\def\ssp{\Omega} 
\def\esf{\mathcal{A}} 

\def\msp{\mathcal{M}} 
\def\nsp{p} 
\def\prodsp{\prod_{k=1}^{\nsp}\msp_{k}} 
\def\dmax{d_{\max}} 
\def\deuc{d_{\mathrm{E}}} 

\def\robj{X} 
\def\vrobj{\bm{\robj}} 
\def\obj{x} 
\def\vobj{\bm{\obj}} 
\def\jtsupp{\supp(\prob\vrobj\inv)} 

\def\prf#1{F^{#1}} 
\def\lbdprf#1{c_{#1}} 
\def\prfdmax{\prf{\vrobj}} 
\def\jprfshort{G}
\def\jprf{\jprfshort^{\vrobj}} 
\def\mprfshort{F}
\def\mprf#1{\mprfshort^{\robj_{#1}}} 
\def\sjprf{\widehat{\jprfshort}^{\vrobj}} 
\def\smprf#1{\widehat{\mprfshort}^{\robj_{#1}}} 
\def\covsjprf{\mc{C}^{\vrobj}} 

\def\rad{r} 
\def\vrad{\bm{\rad}} 
\def\arad{s} 
\def\varad{\bm{\arad}} 
\def\cB{\overline{B}}
\def\cball#1#2#3{\cB_{#1}(#2,#3)} 
\def\doubling{L} 
\def\vone#1{\mathbf{1}_{#1}} 
\def\rdom{\mathcal{D}} 
\def\mrdom{D} 

\def\stat{T} 
\def\w{w} 
\def\sw{\widehat{\w}} 
\def\joint{\mathrm{J}}
\def\prdt{\mathrm{P}}
\def\wjstat{\stat^{\joint,\w}_{n}} 
\def\wpstat{\stat^{\prdt,\w}_{n}}
\def\Ewjstat{\tau^{\joint,\w}} 
\def\Ewpstat{\tau^{\prdt,\w}}
\def\pwjstat#1{\stat^{\joint,\w,#1}_{n}} 
\def\pwpstat#1{\stat^{\prdt,\w,#1}_{n}}
\def\Diff{\Delta^{\vrobj}}
\def\kcwjstat{\overline{\stat}^{\joint,\w}_{\mathrm{c},n}} 
\def\kcwpstat{\overline{\stat}^{\prdt,\w}_{\mathrm{c},n}}

\def\qprfdom{[0,1)} 
\def\vu{\bm{u}} 
 
\def\trvw{(1)} 
\def\covfwjstat{\mc{C}^{\w}} 
\def\dfcovfwjstat{\mc{C}^{\trvw}} 
\def\egnv{\lambda} 
\def\wegnv{\egnv^{\w}}

\def\dfegnv{\egnv^{\trvw}} 
\def\rnorm{Z} 
 
\def\limcdf{\Gamma^{\w}} 
\def\HaEwjstat{H_{a,n}^{\joint,\w}} 
\def\HaEwpstat{H_{a,n}^{\prdt,\w}}
\def\cval{q_\alpha^{\w}} 
\def\pval{\rho} 

\def\sgrp{\mathfrak{S}_n} 
\def\perm{\pi} 
\def\rperm{\Pi} 
\def\nperm{B} 

\def\pcdfwjstat{\wh{\Gamma}^{\joint,\w}_{n,\nperm}} 
\def\pcdfwpstat{\wh{\Gamma}^{\prdt,\w}_{n,\nperm}} 
\def\pcvalwjstat{\wh{q}_{\alpha}^{\joint,\w}} 
\def\pcvalwpstat{\wh{q}_{\alpha}^{\prdt,\w}}
\def\ppvalwjstat{\wh{\pval}_{n,\nperm}^{\joint,\w}} 
\def\ppvalwpstat{\wh{\pval}_{n,\nperm}^{\prdt,\w}}

\def\cw{\constant[\w]} 
\def\Ljprf{L_{\jprfshort}} 

\def\func#1{h_{#1}} 
\def\covrn{N} 
 
\def\brcktn{N_{[]}} 

\def\lcw{c_{\w}} 

\def\testclass{\mathbb{T}_n}
\def\decision{\psi}
\def\tIIerj{R_n^{\joint,\w}} 
\def\tIIerp{R_n^{\prdt,\w}}
\def\haj#1{H_{a,#1}^{\joint,\w}}
\def\hap#1{H_{a,#1}^{\prdt,\w}}

\endlocaldefs

\begin{document}

\begin{frontmatter}
\title{DiPMInd: Distance Profile based\\ Mutual Independence testing for random objects}
\runtitle{DiPMInd}

\begin{aug}
\author[A]{\fnms{Yaqing}~\snm{Chen}\ead[label=e1]{yqchen@stat.rutgers.edu}}\thanksref{t1}
\and
\author[B]{\fnms{Paromita}~\snm{Dubey}\ead[label=e2]{paromita@marshall.usc.edu}}\thanksref{t1}
\thankstext{t1}{Contributed equally to the paper.}
\address[A]{Rutgers University\printead[presep={,\ }]{e1}}

\address[B]{University of Southern California\printead[presep={,\ }]{e2}}
\end{aug}

\begin{abstract}
This paper develops a novel unified framework for testing mutual independence among random objects residing in possibly different metric spaces. The framework generalizes existing methodologies and introduces new measures of mutual independence, and proposes associated tests that achieve minimax rate optimality and exhibit strong empirical power. The foundation of the proposed tests is the new concept of joint distance profiles, which uniquely characterize the joint law of random objects under a mild condition on either the joint law or the metric spaces. Our test statistics quantify the difference of the joint distance profiles of each data point with respect to the joint law and the product of marginal laws of the vector of random objects. To enhance power, we consider integrating this difference with respect to different measures and incorporate flexible data-adaptive weight profiles in the test statistics. We derive the limiting distribution of the test statistics under the null hypothesis of mutual independence and show that the proposed tests with certain weight profiles are asymptotically distribution-free if the marginal distance profiles are continuous. Furthermore, we establish the consistency of the tests under sequences of alternative hypotheses converging to the null. For practical implementations, we employ a permutation scheme to approximate the $p$-values and provide theoretical guarantees that the permutation-based tests maintain type I error control under the null and achieve consistency under alternatives. We demonstrate the power of the proposed tests across various types of data objects through simulations and real data applications, where the new tests exhibit better performance compared with popular existing approaches.
\end{abstract}



\end{frontmatter}


\section{Introduction}\label{sec:intro}
Independence testing is a fundamental problem in statistics, essential for understanding data structures and ensuring reliable inferences, and therefore, is important for both theoretical analysis and practical applications.
First introduced for real-valued random variables as early as by \cite{hoef:48} and \cite{blum:61}, independence testing has a rich history spanning decades and remains a thriving research area, with extensions to random vectors \citep[][among others]{schw:81,gies:97,gret:05,szek:07,hell:13,hell:16,berg:14,zhu:17,pfis:18,weih:18,berr:19,shi:22:2,niu:22,deb:23}, high-dimensional data \citep[][among others]{bisw:16,han:17,drto:20,gao:21,cai:24,zhan:24} and functional data \citep[][among others]{pan:19:2,lai:21,zhu:24} actively explored to date. 

As real-world data have become increasingly complex and often appear in structured forms such as shapes, networks, and probability distributions, metric spaces offer a natural framework for their analysis since  Euclidean coordinates or standard algebraic operations such as additions and scalar multiplications are typically unavailable. Such data, commonly referred to as \textit{random objects}, \textit{object-oriented data}, or \textit{object data}, can be considered as random elements taking values in metric spaces.
With the surge in the availability of object data, this field has emerged as an active research area over the past couple of decades, spurring the development of many new methodologies \citep[see][for reviews on analysis of random objects]{marr:14,mull:16:1,marr:21,dube:24}.

Due to the lack of linear structures, testing mutual independence for random objects is more challenging than that for random vectors and functional data lying in a Hilbert space or a Banach space. While methods such as the distance covariance \citep{szek:07} together with its generalization to mutual independence \citep{jin:18}, the d-variable Hilbert--Schmidt Independence Criterion (dHSIC) \citep{pfis:18}, the sign covariance \citep{berg:14} and the ball covariance \citep{pan:19:2} have been extended to general metric spaces under certain conditions \citep{lyon:13,sejd:13, moon:22, wang:24}, the problem of testing mutual independence among random objects remains underexplored.

In this work, we propose a new unified framework for mutual independence testing based on the characterization of probability measures on metric spaces through distance profiles. 
Introduced in \cite{dube:24}, the distance profile of a point with respect to the law of a random object taking values in a metric space is the distribution of distances between that point and the random object. 
Our approach is inspired by the result presented in Theorem~\ref{thm:characterization} that the joint law of random objects can be characterized entirely by the family of joint distance profiles defined in Eq.~\eqref{eq:joint_prof}, evaluated at all points in its support if it satisfies a mild doubling condition in Eq.~\eqref{eq:limit_doubling}. Hence, mutual independence of a vector of random objects is fully captured by the equality of the joint distance profiles with respect to the joint law and the product of marginal laws, evaluated at all support points of the joint law.

We construct the test statistics by aggregating the weighted integral of the squared differences between the joint distance profiles and the products of the marginal distance profiles across the observations. This yields a class of tests, each associated with different weight profiles and the measures with respect to which the integral is taken. In particular, specific choices of the weight profiles used in the integral with respect to the joint law coincide with the ball covariance test, originally proposed for Banach spaces and later extended to metric spaces \citep{pan:19:2,wang:24}, and the test proposed in \cite{hell:13} as special cases. In the case of testing independence for bivariate random objects, with trivial weights constantly equal to $1$ and the joint law as the measure underlying the integral, one of the proposed test statistics is equivalent to aggregating the Blum--Kiefer--Rosenblatt test statistic \citep{blum:61} over the joint distribution of distances from each observation. 
By replacing multivariate joint distribution functions with joint distance profiles anchored at data points, our work introduces a conceptual and methodological innovation that extends the classical Hoeffding and Blum--Kiefer--Rosenblatt tests in a principled way to random objects residing in potentially different metric spaces. Moreover, the proposed framework offers additional flexibility through the use of different weight profiles and integrating measures, and this flexibility can substantially enhance test power as demonstrated by empirical results. 

While the ball covariance test \citep{pan:19:2,wang:24} and the test proposed in \cite{hell:13} are special cases of one of our test statistics $\wjstat$ under particular choices of weight profiles and integrating measures, 
this paper introduces new methodologies and theoretical results that extend the existing literature. The other class of tests using $\wpstat$ in Eq.~\eqref{eq:test_stat} is novel to the best of our knowledge and demonstrates superior power in simulations. 
Our asymptotic theory comprehensively covers both classes of test statistics, $\wpstat$ and $\wjstat$, and applies to any choice of weights that can be consistently estimated from the data. Using tools from empirical process theory and $U$-process theory, we derive the explicit asymptotic distribution of the proposed test statistics under the null hypothesis of mutual independence. 
In particular, we show that tests with trivial weights are asymptotically distribution-free when the marginal distance profiles are continuous, with the limiting law given by an infinite mixture of independent $\chi^2_1$ random variables with the mixture weights determined by the eigenvalues of a covariance operator (Corollary~\ref{cor:dfree}). 
In the case of testing independence between two random objects, these eigenvalues can be explicitly obtained as the eigenvalues of the product of the covariance operators of standard univariate Brownian bridges, thereby uncovering the same limiting distribution of the Blum–Kiefer–Rosenblatt test statistic \citep{blum:61} for testing independence for bivariate Euclidean data.   
This result extends the literature on distribution-free independence tests developed for random vectors \citep[e.g.,][]{han:17,niu:22, deb:23,shi:22:2} and high dimensional data \citep{cai:24} to random objects---a new result that has remained elusive in existing approaches within metric spaces and is of independent interest. 
To the best of our knowledge, these are the first asymptotically distribution-free tests for mutual independence for general metric-space-valued random objects. 
The proposed testing framework also incorporates flexible, data-adaptive weight profiles, with certain choices shown empirically to deliver strong power. For practical implementations, we employ a permutation scheme to compute $p$-values,
and establish its validity both under the null of mutual independence and under sequences of alternatives converging to the null, thereby providing guarantees lacking in existing methods. 
Moreover, we establish that the proposed tests attain the minimax rate optimality. 
The asymptotic theory established for the proposed tests supports a broad range of choices beyond the limited set of weight profiles and integrating measures that have been previously considered. This flexibility not only enables optimization or ensembling across multiple choices to enhance power, but also provides a unified and principled framework for mutual independence testing in general metric spaces.

We evaluate the empirical performance of our omnibus tests by benchmarking them against strong existing competitors, specifically the distance covariance and the dHSIC, through extensive simulations across both Euclidean and non-Euclidean settings. The tests within the proposed framework, especially the new tests, outperform the other tests in many scenarios, e.g., when data distributions are heavy-tailed or corrupted by outliers. We also apply our tests to real data to examine for each season the independence between hourly bike rental compositions and the distributions of temperature, humidity, and wind speed, where the compositional data and all the distributional data are jointly recorded for each day, with multiple comparison adjustments across seasons. The proposed tests are capable of detecting dependence relationships that are missed by competing methods.

In Section~\ref{sec:method}, we introduce the notion of joint distance profiles, establish their unique characterization property, and propose tests for mutual independence of random objects based on comparing joint distance profiles with respect to the joint law of the vector of random objects of interest and the product of marginal laws, which, under the null of mutual independence, is equal to the joint law. 
In addition, we present the theoretical guarantees of our tests, ensuring type I error rate control under the null and the power guarantees under sequences of alternatives converging towards the null for both the asymptotic tests and permutation-based tests and establishing the minimax rate optimality. 
Auxiliary theoretical results and proofs are provided in Section~S.1 
in the Supplement. 
Section~\ref{sec:simulations} details our empirical experiments, followed by real-world data applications in Section~\ref{sec:data}.

\section{Methods and theory}\label{sec:method}

Let $(\msp_1,d_1),\dots,(\msp_{\nsp},d_{\nsp})$ be $\nsp$ separable metric spaces and $\borel(\msp_k)$ be the corresponding Borel $\sigma$-algebras for $k=1,\dots,\nsp$. 
For the product space $\prodsp$, consider a metric $\dmax$ given by $\dmax(\vobj,\vobj') = \max_{1\le k\le \nsp}\{d_k(\obj_{k},\obj'_{k})\}$, for any $\vobj = (\obj_1,\dots,\obj_{\nsp})\tps$ and $\vobj' = (\obj'_1,\dots,\obj'_{\nsp})\tps$ in $\prodsp$. 
The Borel $\sigma$-algebra of the metric space $(\prodsp, \dmax)$ is given by $\borel(\prodsp) = \borel(\msp_{1})\times \dots\times \borel(\msp_{\nsp})$. 
A multivariate random object $\vrobj = (\robj_1,\dots,\robj_{\nsp})\tps$ taking values in $\prodsp$ is a measurable map from an underlying probability space $(\ssp,\esf,\prob)$ to $(\prodsp, \borel(\prodsp))$, where within the tuple $(\ssp,\esf,\prob)$, $\ssp$ is the sample space of all possible outcomes, $\esf$ is a $\sigma$-algebra of events, which are subsets of $\ssp$, and $\prob$ is a probability measure. 
Testing whether $\robj_1,\dots,\robj_{\nsp}$ are mutually independent requires the comparison between the joint law of $\vrobj$, $\prob\vrobj\inv$, and the product measure of the marginals, $\prob \robj_1\inv \times \dots\times \prob \robj_{\nsp}\inv$. In particular, it concerns the null and alternative hypotheses given by 
\bal\label{eq:hypo}
    &H_0:\ \prob \vrobj\inv = \prob \robj_1\inv \times \dots\times \prob \robj_{\nsp}\inv\\
    \text{versus } &H_a:\ \prob \vrobj\inv \ne \prob \robj_1\inv \times \dots\times \prob \robj_{\nsp}\inv,
\eal
respectively. 
We propose \textit{joint distance profiles} to quantify the joint law of random objects $\robj_1,\dots,\robj_{\nsp}$. 
Specifically, for any given $\vobj = (\obj_1,\dots,\obj_{\nsp})\tps\in\prodsp$, the \emph{joint distance profile} $\jprf_{\vobj}$ of $\vobj$ with respect to the joint law of $\vrobj$ is defined as a function that maps $\real^{\nsp}$ to $[0,1]$ with  
\bal\label{eq:joint_prof}
    \jprf_{\vobj} (\vrad) = \prob\{d_1(\obj_1,\robj_1)\le \rad_1,\dots, d_{\nsp}(\obj_{\nsp},\robj_{\nsp})\le \rad_{\nsp}\}
\eal 
for $\vrad=(\rad_1,\dots,\rad_{\nsp})\tps\in\real^{\nsp}$. 
Here and later, we suppress the dependence on the metrics in the notation where it is not ambiguous. 

\begin{remark}
When $\nsp=1$, the notion of joint distance profiles coincides with that of distance profiles proposed by \cite{dube:24} for characterizing the law of random objects taking values in metric spaces. 
However, when $\nsp>1$, the two concepts diverge and present alternative viewpoints---using joint distance profiles views $\vrobj$ as a vector of random objects each lying in a metric space $(\msp_{k},d_{k})$, as opposed to using distance profiles by \cite{dube:24}, which views $\vrobj$ as a single random object in the product space $\prodsp$ endowed with a metric, e.g, $\dmax$ or the $q$-product metric $d_{q} = (\sum_{k=1}^{\nsp}d_{k}^{q})^{1/q}$ with some $q\in[1,+\infty)$. 
In addition, while the distance profiles proposed by \cite{dube:24} can be derived from joint distance profiles for any metric that metrizes the product space and depends only on the metrics $d_1,\dots,d_{\nsp}$, the converse does not generally hold. 
In particular, considering the product metric space $(\prodsp,\dmax)$, the \textit{distance profile} $\prfdmax_{\vobj}$ of $\vobj\in\prodsp$ with respect to $\prob\vrobj\inv$ is a function from $\real$ to $[0,1]$, which can be expressed as  
\bal\nn
    \prfdmax_{\vobj}(\rad) 
    = \prob\{\dmax(\vobj,\vrobj)\le \rad\} 
    = \prob\{d_1(\obj_1,\robj_1)\le \rad,\dots, d_{\nsp}(\obj_{\nsp},\robj_{\nsp})\le \rad\} 
    = \jprf_{\vobj}(\rad\vone{\nsp})
\eal
for $\rad\in\real$, where $\vone{\nsp}$ denotes a $\nsp$-dimensional vector with all entries $1$. 
Thus, in this case, the distance profile $\prfdmax_{\vobj}$ recovers the joint distance profile $\jprf_{\vobj}$ only along the diagonal $\{\vrad\in\real^{\nsp}: \rad_1=\rad_2=\dots=\rad_{\nsp}\}$ in addition to regions where definitions dictate zero probability $\{\vrad\in\real^{\nsp}: \exists k\text{ such that } \rad_k<0\}$. 
Consequently, when $\nsp>1$, joint distance profiles are strictly more informative than distance profiles. 
\end{remark}

As the simplest marginals of the stochastic processes $\{\dmax(\vobj,\vrobj):\, \vobj\in\prodsp\}$ and $\{( d_1(\obj_1,\robj_1), \dots, d_{\nsp}(\obj_{\nsp},\robj_{\nsp}) )\tps:\,\vobj\in\prodsp\}$, respectively, 
distance profiles $\prfdmax_{\vobj}$ and joint distance profiles $\jprf_{\vobj}$ are the mass assigned by $\prob\vrobj\inv$ of closed balls $\cball{\dmax}{\vobj}{\rad}$ and products of marginal closed balls $\prod_{k=1}^{\nsp}\cball{d_k}{\obj_k}{\rad_k}$, respectively. 
Here, for any metric space $(\msp,d)$, $\cball{d}{\obj}{\rad}$ denotes a closed ball centering at $\obj\in\msp$ with radius $\rad>0$ in terms of the metric $d$, i.e., $\cball{d}{\obj}{\rad} = \{\obj'\in\msp: d(\obj,\obj') \le \rad\}$. 
A question of interest is whether, and under what conditions, $\prob\vrobj\inv$ can be uniquely determined by the mass it assigns to closed balls or to products of closed balls. 
In Theorem~\ref{thm:characterization}, we show that a sufficient condition for such unique characterization is that the probability measures belong to a family of measures $\mu$ such that with some $\doubling_{\mu} > 0$, 
\bal\label{eq:limit_doubling}
    \lim_{\rad\downarrow 0} \frac{\mu\{\cball{d}{\obj}{2\rad}\}}{\mu\{\cball{d}{\obj}{\rad}\}} \le \doubling_{\mu}  \text{ for all } \obj\in\supp(\mu),
\eal
where $\supp(\mu)$ denotes the support of $\mu$, which is the closure of the set of all points of which $\mu$ assigns positive mass to every open neighborhood, i.e., 
\bal\nn
    \supp(\mu) = \mathrm{cl}\left( \{ \obj\in\msp: \mu(U)>0 \text{ for all open set } U\ni \obj\} \right).
\eal 
The condition in Eq.~\eqref{eq:limit_doubling} is a weaker variant of the doubling condition, the latter requires the ratio in Eq.~\eqref{eq:limit_doubling} to be uniformly bounded for all $r>0$. 
In Remark~\ref{rmk:doubling} below, we provide interpretable sufficient conditions, along with examples of probability measures, that satisfy Eq.~\eqref{eq:limit_doubling}. 

\begin{remark}\label{rmk:doubling}
For $(\msp,d)$ with a reference measure $\nu$ that is a doubling measure in the sense that $\nu\{\cball{d}{\obj}{2\rad}\}\le \doubling'_{\nu}\nu\{\cball{d}{\obj}{\rad}\}$ for all $\rad>0$ and $\obj\in\msp$ for some $\doubling'_{\nu}>0$, 
if a probability measure $\mu$ on $\msp$ is absolutely continuous with respect to $\nu$ and the Radon--Nikodym derivative $\frac{\diffop\mu}{\diffop\nu}$ is positive and continuous over $\supp(\mu)$, then $\mu$ satisfies Eq.~\eqref{eq:limit_doubling}. 
In the special case of $\msp=\real^{\nsp}$ endowed with the Euclidean metric $\deuc$, this means that any probability measure on $\real^{\nsp}$ with continuous positive density function satisfies Eq.~\eqref{eq:limit_doubling}. 
Another sufficient condition of a measure $\mu$ satisfying Eq.~\eqref{eq:limit_doubling} is an Ahlfors--David type assumption \citep{ahlf:35,davi:85}: There exist constants $s>0$, $C_{\mu}>1$ and $c_{\mu}>0$ such that 
\bal\nn
C_{\mu}\inv\rad^s \le \mu\{\cball{d}{\obj}{\rad}\} \le C_{\mu}\rad^s
\ \text{for all } \obj\in\supp(\mu) \text{ and }\rad\in(0,c_{\mu}]. 
\eal 
On the other hand, the condition in Eq.~\eqref{eq:limit_doubling} does not necessarily impose such smoothness requirements on the distance profiles with respect to $\mu$, $\prf{\mu}_{\obj}\colon [0,+\infty)\ra [0,1]$ with $\rad\mapsto \mu\{\cball{d}{\obj}{\rad}\}$ for $\obj\in\supp(\mu)$.
Note that the condition in Eq.~\eqref{eq:limit_doubling} only involves balls with bounded radii centered in the support of the corresponding probability measure instead of all the closed balls in the metric space. 
Besides Ahlfors--David type measures, 
a discrete measure $\mu$ such that $\mu(\{\obj\})\ge \lbdprf{\mu}'$ for all $\obj\in\supp(\mu)$ for some $\lbdprf{\mu}'>0$ also satisfies Eq.~\eqref{eq:limit_doubling}. 
We refer to \cite{hein:01} for further discussion on the doubling condition. 
\end{remark}

\begin{theorem}\label{thm:characterization}
Let $(\msp,d)$ be a separable metric space and let $\mu_1$ and $\mu_2$ be two probability measures on $(\msp,\borel(\msp))$ satisfying the condition in Eq.~\eqref{eq:limit_doubling}. 
Then $\mu_1$ and $\mu_2$ agree if and only if there exists a Borel set $S_1\subseteq\supp(\mu_1)$ such that $\mu_1(S_1)=1$ and that $\mu_1$ and $\mu_2$ agree on all closed balls in $\borel(\msp)$ with centers in $S_1$, i.e., $\mu_1\{\cball{d}{\obj}{\rad}\}  = \mu_2\{\cball{d}{\obj}{\rad}\}$ for all $\obj\in S_1$ and $\rad>0$. 
\end{theorem}

The unique characterization of probability measures on a separable metric space through the mass assigned to closed balls has been discussed in Proposition~1 of \cite{dube:24}; specifically, if $(\msp,d^u)$ is of strong negative type \citep{kleb:05,lyon:13} for some $u>0$, then two probability measures $\mu_1$ and $\mu_2$ on $\msp$ are equal if they assign the same mass to every closed ball centered in the union of their supports. 
The condition given by \cite{dube:24} concerns the metric space, whereas the condition in Eq.~\eqref{eq:limit_doubling} is regarding the probability measures on the metric space, 
so the two conditions can complement each other as guarantees of the unique characterization of probability measures using closed balls. 
In addition, we note that under the condition in Eq.~\eqref{eq:limit_doubling}, two probability measures coincide as long as they are equal on all closed balls centered in  \textit{either} of their supports, in contrast to the case under the condition on the metric space given by \cite{dube:24}, the equality of two probability measures is guaranteed by the equality of mass on every closed ball centered in \textit{both} of their supports. 

For any $k=1,\dots,\nsp$ and $\obj_{k}\in\msp_{k}$, let $\mprf{k}_{\obj_{k}}$ be the distance profile of $\obj_{k}$ with respect to the marginal law of $\robj_{k}$, respectively, i.e., 
\bal
    \mprf{k}_{\obj_{k}}(\rad) = \prob\{ d_k(\obj_{k},\robj_{k}) \le \rad \}
    \label{eq:marg_prof}
\eal
for $\rad\in\real$. 
Consider the following three cases:
\begin{enumerate}[label = (C\arabic*)]
    \item\label{ass:doubling_joint} The joint law $\prob\vrobj\inv$ satisfies the condition in Eq.~\eqref{eq:limit_doubling}.
    \item\label{ass:doubling_product} The product law $\prob \robj_1\inv \times \dots\times \prob \robj_{\nsp}\inv$ satisfies the condition in Eq.~\eqref{eq:limit_doubling}.
    \item\label{ass:doubling_marginal} Each marginal law $\prob\robj_k\inv$ for $k=1,\dots,\nsp$ satisfies the condition in Eq.~\eqref{eq:limit_doubling}. 
\end{enumerate}
Suppose that one of the three conditions \ref{ass:doubling_joint}--\ref{ass:doubling_marginal} holds. 
Then by Theorem~\ref{thm:characterization}, 
$\prob \vrobj\inv = \prob \robj_1\inv \times \dots\times \prob \robj_{\nsp}\inv$ holds if and only if for all $\vobj\in\supp(\prob\vrobj\inv)$ and $\rad > 0$, 
\bal\nn
\prob\vrobj\inv\{\cball{\dmax}{\vobj}{\rad}\} = (\prob \robj_1\inv \times \dots\times \prob \robj_{\nsp}\inv) \{\cball{\dmax}{\vobj}{\rad}\}. 
\eal 
That is, $\robj_1,\dots,\robj_{\nsp}$ are mutually independent if and only if for all $\vobj\in\supp(\prob\vrobj\inv)$ and $\rad > 0$, 
\bal\nn
\prob\vrobj\inv\left\{\prod_{k=1}^{\nsp}\cball{d_k}{\obj_k}{\rad}\right\} = \prod_{k=1}^{\nsp}\prob \robj_k\inv \{\cball{d_k}{\obj_k}{\rad}\}. 
\eal
Therefore, a sufficient and necessary condition for the mutual independence of 
$\robj_{1},\dots,\robj_{\nsp}$, i.e., $H_0$ in Eq.~\eqref{eq:hypo}, is that for all $\vobj\in\supp(\prob\vrobj\inv)$ and $\vrad \in\real^{\nsp}$, 
\bal\label{eq:indep_jprf}
    \jprf_{\vobj}(\vrad) = \prod_{k=1}^{\nsp} \mprf{k}_{\obj_{k}}(\rad_k). 
\eal
This motivates defining measures of mutual independence by quantifying the difference between the two sides of Eq.~\eqref{eq:indep_jprf}.  
Consider an independent copy of $\vrobj$, $\vrobj'\sim\prob\vrobj\inv$. 
As per Eq.~\eqref{eq:joint_prof}, the joint distance profile $\jprf_{\vrobj'}$ of $\vrobj'$ with respect to $\prob\vrobj\inv$ is 
\bal\nn
    \jprf_{\vrobj'}(\vrad)
    = \prob\{ d_1(\robj'_{1},\robj_1)\le \rad_1,\dots, d_{\nsp}(\robj'_{\nsp},\robj_{\nsp})\le \rad_{\nsp} \cond \vrobj'\}
\eal
for $\vrad\in\real^{\nsp}$. 
From Eq.~\eqref{eq:marg_prof}, for any $k=1,\dots,\nsp$, the distance profile $\mprf{k}_{\robj'_{k}}$ of $\robj'_{k}$ with respect to $\prob\robj_{k}\inv$ is 
\bal\nn
    \mprf{k}_{\robj'_{k}}(\rad) = \prob\{ d_k(\robj'_{k},\robj_{k}) \le \rad \cond \robj'_{k} \}
\eal
for $\rad\in\real$. 
To construct measures of mutual independence, we furthermore introduce a weight profile $\w_{\vobj}\colon \real^{\nsp}\ra [0,+\infty)$ for each $\vobj\in\prodsp$ such that $\w_{\vobj}$ is positive $\prod_{k=1}^{\nsp}\mprf{k}_{\obj_k}$-almost everywhere. A trivial choice is $\w_{\vobj}\equiv 1$ for all $\vobj\in\prodsp$. 
We define two quantities:    
\bal\label{eq:Ewjstat}
\Ewpstat &= \E\left(\int \w_{\vrobj'}(\vrad) \left\{\jprf_{\vrobj'}(\vrad) 
-  \prod_{k=1}^{\nsp}\mprf{k}_{\robj'_{k}}(\rad_k)\right\}^2 \prod_{k=1}^{\nsp}\diffop\mprf{k}_{\robj'_{k}}(\rad_{k})\right)\\
\text{and}\quad 
\Ewjstat &= \E\left(\int \w_{\vrobj'}(\vrad) \left\{\jprf_{\vrobj'}(\vrad) 
-  \prod_{k=1}^{\nsp}\mprf{k}_{\robj'_{k}}(\rad_k)\right\}^2\diffop\jprf_{\vrobj'}(\vrad)\right). 
\eal 
By Theorem~\ref{thm:characterization}, $\Ewpstat$ and $\Ewjstat$ serve as measures of mutual independence among $\robj_1,\dots,\robj_{\nsp}$ (Proposition~\ref{prop:indep_measure} below). 
A related work \citep{zhou:25v1}, posted on arXiv after the initial version of our manuscript \citep{chen:24v1}, studies measures of association and independence tests based on the special case of $\Ewjstat$ with trivial weights and $\nsp=2$. 

\begin{proposition}
\label{prop:indep_measure}
(i) Assume that either the joint law $\prob\vrobj\inv$ or each of the marginals $\prob\robj_{k}\inv$ (for $k=1,\dots,\nsp$) satisfies the condition in Eq.~\eqref{eq:limit_doubling}. Then, the following hold:
\begin{itemize}
    \item $\prob\vrobj\inv = \prob\robj_1\inv\times\dots\times\prob\robj_{\nsp}\inv$ if and only if $\Ewpstat = 0$. 
    \item If, additionally, the product measure is absolutely continuous with respect to the joint law (i.e., $\prob\robj_1\inv\times\dots\times\prob\robj_{\nsp}\inv \ll \prob\vrobj\inv$), then $\prob\vrobj\inv = \prob\robj_1\inv\times\dots\times\prob\robj_{\nsp}\inv$ if and only if $\Ewjstat = 0$.
\end{itemize}
(ii) If $(\msp_k,d_k)$ is of strong negative type for each $k=1,\dots,\nsp$ and $\prob\robj_1\inv\times\dots\times\prob\robj_{\nsp}\inv \ll \prob\vrobj\inv$, then the following statements are equivalent:
\begin{itemize}
    \item $\prob\vrobj\inv = \prob\robj_1\inv\times\dots\times\prob\robj_{\nsp}\inv$. 
    \item $\Ewpstat = 0$. 
    \item $\Ewjstat = 0$.
\end{itemize}
\end{proposition}

This motivates testing the mutual independence of $\robj_1,\dots,\robj_{\nsp}$ based on estimates of $\Ewpstat$ and $\Ewjstat$ from a sample of $n$ observations $\vrobj_{1},\dots,\vrobj_{n}$ 
such that each $\vrobj_{i} = (\robj_{i,1},\dots,\robj_{i,\nsp})\tps \sim \prob\vrobj\inv$ for $i=1,\dots,n$ 
and that $\vrobj,\vrobj',\vrobj_{1},\vrobj_{2},\dots,\vrobj_{n}$ are mutually independent. 

For any $\vobj\in\prodsp$, the sample estimates of its joint distance profile and marginal distance profiles are given by 
\bal\label{eq:sprf}
    \sjprf_{\vobj}(\vrad) 
    &= \frac{1}{n}\sum_{j=1}^{n} \indicator\{d_1(\obj_{1},\robj_{j,1})\le \rad_1,\dots, d_{\nsp}(\obj_{\nsp},\robj_{j,\nsp})\le \rad_{\nsp}\}\\
    \text{and } \smprf{k}_{\obj_{k}}(\rad)
    &= \frac{1}{n}\sum_{j=1}^{n} \indicator\{d_k(\obj_{k},\robj_{j,k})\le \rad\},
\eal
respectively, for $\vrad\in\real^{\nsp}$ and $\rad\in\real$. 
Furthermore, for each observation $\vrobj_{i}$, the sample estimates $\sjprf_{\vrobj_{i}}$ and $\{\smprf{k}_{\robj_{i,k}}\}_{k=1}^{\nsp}$ of its joint distance profile and marginal distance profiles are obtained by plugging $\vobj=\vrobj_{i}$ into Eq.~\eqref{eq:sprf}. 
Let $\sw_{\vobj}\colon \real^{\nsp}\ra [0,+\infty)$ be the sample estimate of the weight profile $\w_{\vobj}$ for each $\vobj\in\prodsp$. 
Motivated by $\Ewpstat$ and $\Ewjstat$ in Eq.~\eqref{eq:Ewjstat}, we define the proposed test statistics as  
\bal\label{eq:test_stat}
\wpstat(\vrobj_1,\vrobj_2,\dots,\vrobj_n) 
&= \sum_{i=1}^{n} \int \sw_{\vrobj_{i}}(\vrad) \left\{ \sjprf_{\vrobj_{i}}(\vrad) - \prod_{k=1}^{\nsp} \smprf{k}_{\robj_{i,k}}(\rad_{k}) \right\}^2 \prod_{k=1}^{\nsp}\diffop\smprf{k}_{\robj_{i,k}}(\rad_{k})\\
\text{and}\ 
\wjstat(\vrobj_1,\vrobj_2,\dots,\vrobj_n) 
&= \sum_{i=1}^{n} \int \sw_{\vrobj_{i}}(\vrad) \left\{ \sjprf_{\vrobj_{i}}(\vrad) - \prod_{k=1}^{\nsp} \smprf{k}_{\robj_{i,k}}(\rad_{k}) \right\}^2 \diffop\sjprf_{\vrobj_{i}}(\vrad),
\eal
respectively. 
For simplicity, we suppress the dependence on the data in the notation and denote $\wpstat = \wpstat(\vrobj_1,\vrobj_2,\dots,\vrobj_n)$ and $\wjstat = \wjstat(\vrobj_1,\vrobj_2,\dots,\vrobj_n)$ wherever unambiguous. 

For the proposed tests, we establish theoretical guarantees for type I error control under the null hypothesis and analyze their power under alternative hypotheses.  
We begin by investigating the asymptotic properties of the proposed test statistics in Eq.~\eqref{eq:test_stat} under $H_0$. 
This analysis requires a thorough understanding of the asymptotic behavior of the sample estimates of joint distance profiles $\sjprf_{\vobj}(\cdot)$, as defined in Eq.~\eqref{eq:sprf}, which serves as the foundation for our approach. 
For each $k=1,\dots,\nsp$, let $\diam(\msp_{k}) = \sup\{d_k(\obj_k,\obj'_k):\, \obj_k,\obj'_k\in\msp_k\}$, which may be finite or infinite. 
We define $\rdom = \prod_{k=1}^{\nsp}\mrdom_k$, where $\mrdom_k = [0,\diam(\msp_{k})]$ if $\diam(\msp_{k})$ is finite and $\mrdom_k = [0,+\infty)$ otherwise. 
For $\vobj\in\prodsp$ and $\vrad\in\rdom$, we define the function $\function[\vobj,\vrad]{}\label{fcl:sjprf}\colon\prodsp\ra\real$ given by  
\bal\label{eq:func_sjprf}
\function[\vobj,\vrad][\ref{fcl:sjprf}](\vobj') 
=\indicator\left\{\vobj'\in\prod_{k=1}^{\nsp}\cball{d_k}{\obj_k}{\rad_k}\right\}
= \indicator\{d_1(\obj_1,\obj'_1)\le\rad_1,\dots,d_{\nsp}(\obj_{\nsp},\obj'_{\nsp})\le\rad_{\nsp}\}
\eal
for $\vobj'\in\prodsp$. Let 
\bal\label{eq:fcl_sjprf}
\fclass[\ref{fcl:sjprf}]= \left\{\function[\vobj,\vrad][\ref{fcl:sjprf}]:\, \vobj\in\prodsp,\, \vrad\in\rdom\right\}
\eal
denote the collection of all such functions. 
The function class $\fclass[\ref{fcl:sjprf}]$ determines 
the joint distance profiles $\jprf_{\vobj}(\cdot)$ and their sample estimates $\sjprf_{\vobj}(\cdot)$ defined in Eqs.~\eqref{eq:joint_prof} and \eqref{eq:sprf}, respectively; specifically, 
\bal\nn
\jprf_{\vobj}(\vrad) = \E\left\{\function[\vobj,\vrad][\ref{fcl:sjprf}](\vrobj)\right\} \quad\text{and}\quad 
\sjprf_{\vobj}(\vrad) = \frac{1}{n}\sum_{i=1}^{n} \function[\vobj,\vrad][\ref{fcl:sjprf}](\vrobj_{i})
\eal
for all $\vrad\in\rdom$ and $\vobj\in\prodsp$. 
As a crucial step in analyzing the theoretical properties of the proposed methods, 
we establish the Donsker property of the function class $\fclass[\ref{fcl:sjprf}]$ in Theorem~\ref{thm:sjprl}, 
which relies on an assumption on the bracketing number of $\fclass[\ref{fcl:sjprf}]$. 
For any $\eps>0$, an $\eps$-bracket in terms of $L_1(\prob\vrobj\inv)$ norm is a bracket $[l,u]$ determined by two functions $l,u\colon\prodsp\ra\real$, which consists of all functions $\func{}\colon\prodsp\ra\real$ satisfying $l\le \func{} \le u$ and $\lVert l-u\rVert_{L_1(\prob\vrobj\inv)} \coloneqq \E\lvert l(\vrobj)-u(\vrobj)\rvert <\eps$. 
We denote the bracketing number by $\brcktn\{\eps,\fclass[\ref{fcl:sjprf}],L_1(\prob\vrobj\inv)\}$, i.e., the minimum number of such $\eps$-brackets needed to cover $\fclass[\ref{fcl:sjprf}]$, for which we assume 
\begin{assumption}\label{ass:fcl_sjprf_brcktn} 
$\eps\log\brcktn\{\eps,\fclass[\ref{fcl:sjprf}],L_1(\prob\vrobj\inv)\}\ra 0$ as $\eps\downarrow 0$. 
\end{assumption}
Under Assumption~\ref{ass:fcl_sjprf_brcktn}, we establish that the function class $\fclass[\ref{fcl:sjprf}]$ is $\prob\vrobj\inv$-Donsker:

\begin{theorem}\label{thm:sjprl}
Under Assumption~\ref{ass:fcl_sjprf_brcktn}, the process $\{\sqrt{n}\{\sjprf_{\vobj}(\vrad)-\jprf_{\vobj}(\vrad)\}:\,\vobj\in\prodsp,\,\newline \vrad\in\rdom\}$ converges weakly to a Gaussian process
with mean zero and covariance $\covsjprf\colon \newline (\prodsp\times\rdom)^2\ra\real$, as $n\ra\infty$, where 
\bal\nn 
&\covsjprf(\vobj,\vrad,\vobj',\vrad') 
= \prob\left(d_k(\obj_k,\robj_{i,k})\le\rad_k,\,d_k(\obj'_k,\robj'_{i,k})\le\rad'_k,\, \forall k=1,\dots,\nsp\right) - \jprf_{\vobj}(\vrad)\jprf_{\vobj'}(\vrad')\eal
for $\vobj,\vobj'\in\prodsp$ and $\vrad,\vrad'\in\rdom$. 
\end{theorem}
 
Assumption~\ref{ass:fcl_sjprf_brcktn} is needed to address the challenges raised by the discontinuity of the indicator functions in $\fclass[\ref{fcl:sjprf}]$ defined in Eq.~\eqref{eq:fcl_sjprf}. 
Similar challenges are also encountered by \cite{dube:24} to establish the Donsker result for sample distance profiles, which is a special case of that for sample joint distance profiles in Theorem~\ref{thm:sjprl} with $\nsp=1$. 
The corresponding assumptions in \cite{dube:24}, adapted to the scenario of estimating joint distance profiles, are: 
\begin{enumerate}[label = (D\arabic*), series = fcl_sjprf_brcktn]
\item \label{ass:msp_covrn}
For each $k=1,\dots,\nsp$, 
$\eps\log \covrn(\eps,\msp_k,d_k)\ra 0$ as $\eps\downarrow 0$, 
where $\covrn(\eps,\msp_k,d_k)$ is the covering number of the space $\msp_k$ with balls of radius $\eps$. 
\item \label{ass:jprfl_lips} 
There exists a constant $\Ljprf>0$ such that $\sup_{\vobj\in\prodsp} \lvert\jprf_{\vobj}(\vrad) - \jprf_{\vobj}(\varad)\rvert \le\Ljprf\lVert\vrad-\varad\rVert$ for all $\vrad,\varad\in\rdom$. 
\end{enumerate}
Assumption~\ref{ass:fcl_sjprf_brcktn} is regarding the bracketing entropy of the function class $\fclass[\ref{fcl:sjprf}]$ that determines the joint distance profiles, 
whereas \ref{ass:msp_covrn} entails that the metric spaces $\{\msp_{k}\}_{k=1}^{\nsp}$ are (totally) bounded and  
\ref{ass:jprfl_lips} requires the joint distance profiles to be (uniformly) Lipschitz continuous, which is violated if $\prob\vrobj\inv$ has atoms. 
We establish in Proposition~\ref{prop:fcl_sjprf_brcktn} that Assumption~\ref{ass:fcl_sjprf_brcktn} is a relaxation of \ref{ass:msp_covrn} and \ref{ass:jprfl_lips}. 
\begin{proposition} \label{prop:fcl_sjprf_brcktn}
Assumption~\ref{ass:fcl_sjprf_brcktn} holds if \ref{ass:msp_covrn} and \ref{ass:jprfl_lips} are both satisfied.
\end{proposition}

\begin{remark}\label{rmk:brcktn}
As discussed in \cite{dube:24}, there are various spaces commonly appearing in real-world data, which satisfy \ref{ass:msp_covrn} and hence also Assumption~\ref{ass:fcl_sjprf_brcktn} if \ref{ass:jprfl_lips} is also satisfied. Examples include: all spaces being subsets of finite dimensional Euclidean spaces, 
such as the space of networks with a fixed number of nodes represented by graph Laplacians or adjacency matrices, spaces of symmetric positive definite matrices of a fixed dimension, spaces of phylogenetic trees with a common number of tips; 
VC-classes of sets or functions; 
2-Wasserstein space of one-dimensional distributions with density functions uniformly bounded from above and away from zero; among many others. 
On the other hand, in contrast to the spaces satisfying \ref{ass:msp_covrn} which are therefore totally bounded, Assumption~\ref{ass:fcl_sjprf_brcktn} does not necessarily require the spaces $\{(\msp_k,d_k)\}_{k=1}^{\nsp}$ to be totally bounded. 
For spaces that are not totally bounded or even unbounded, Assumption~\ref{ass:fcl_sjprf_brcktn} can still hold if $\prob\vrobj\inv$ assigns sufficiently small mass to the complement of any compact set that satisfies \ref{ass:msp_covrn}. 
Specifically, for any $\eps>0$, there exists a closed ball $\cB_k(\rad_\eps)$ of radius $\rad_\eps$ in $\msp_k$ for each $k=1,\dots,\nsp$ such that $\prob\{\vrobj\notin \prod_{k=1}^{\nsp}\cB_k(\rad_\eps)\}\le\eps$. 
Assumption~\ref{ass:fcl_sjprf_brcktn} is satisfied if $\eps\log N\{\eps,\cB_k(\rad_\eps), d_k\} \ra 0$ as $\eps \downarrow 0$ for all $k=1,\dots,\nsp$ in conjunction with \ref{ass:jprfl_lips}. 
For instance, in the case of Euclidean spaces, if each marginal distribution $\prob\robj_k\inv$ has tail decaying at an exponential rate for $k=1,\dots,\nsp$---e.g., if they are subexponential and sub-Gaussian---then Assumption~\ref{ass:fcl_sjprf_brcktn} is satisfied.
\end{remark}

While Theorem~\ref{thm:sjprl} presents the Donsker result for the sample joint distance profiles of all points in the entire space $\prodsp$, the analysis of asymptotic behavior of the proposed test statistics in Eq.~\eqref{eq:test_stat} only concerns the sample joint distance profiles of points in the support of the joint law, $\jtsupp$, which are determined by a subclass of functions  
\bal\label{eq:fcl_sjprf_on_jtsupp}
\fclass[\ref{fcl:sjprf}]' = \left\{\function[\vobj,\vrad][\ref{fcl:sjprf}]:\, \vobj\in\jtsupp,\, \vrad\in\rdom\right\}.
\eal
Moreover, the subclass $\fclass[\ref{fcl:sjprf}]'$ is $\prob\vrobj\inv$-Donsker  
under a weaker version of Assumption~\ref{ass:fcl_sjprf_brcktn}: 
\begin{assumptionp}{\ref{ass:fcl_sjprf_brcktn}$'$}\label{ass:fcl_sjprf_brcktn_prime}
$\eps\log\brcktn\{\eps,\fclass[\ref{fcl:sjprf}]',L_1(\prob\vrobj\inv)\}\ra 0$ as $\eps\downarrow 0$. 
\end{assumptionp}
Under Assumption~\ref{ass:fcl_sjprf_brcktn_prime}, 
the process $\{\sqrt{n}\{\sjprf_{\vobj}(\vrad)-\jprf_{\vobj}(\vrad)\}:\,\vobj\in\jtsupp,\,\vrad\in\rdom\}$ converges weakly to a Gaussian process
with mean zero and covariance $\covsjprf$ defined in Theorem~\ref{thm:sjprl} restricted to $(\jtsupp\times\rdom)^2$. 
With Assumption~\ref{ass:fcl_sjprf_brcktn_prime}, the Donsker property of the subclass $\fclass[\ref{fcl:sjprf}]'$ holds not only when the joint distance profiles are continuous, e.g., when \ref{ass:jprfl_lips} holds, but also when $\prob\vrobj\inv$ is a discrete measure with finite support, in which case \ref{ass:jprfl_lips} is violated.

Next, we establish the asymptotic
distributions of the proposed test statistics in Eq.~\eqref{eq:test_stat} under $H_0$ in Eq.~\eqref{eq:hypo}, for which we require an additional assumption. 

\begin{assumption}\label{ass:weight}
There exists $\cw>0$ such that $\sup_{\vobj\in\prodsp}\sup_{\vrad\in\rdom}\lvert\w_{\vobj}(\vrad)\rvert\le \cw$. 
In addition, the sample estimate of the weight profiles $\sw_{\vobj}$ satisfies that $\frac{1}{n}\sum_{i=1}^{n} \lvert\sw_{\vrobj_{i}}(\vrad)-\w_{\vrobj_{i}}(\vrad)\rvert \diffop\sjprf_{\vrobj_{i}}(\vrad)$ and $\frac{1}{n}\sum_{i=1}^{n} \lvert\sw_{\vrobj_{i}}(\vrad)-\w_{\vrobj_{i}}(\vrad)\rvert \prod_{k=1}^{\nsp}\diffop\smprf{k}_{\robj_{i,k}}(\rad_{k})$ both converge to $0$ in probability as $n\ra\infty$. 
\end{assumption}

Assumption~\ref{ass:weight} is a regularity condition on the weight profiles and their sample estimates 
and is needed to ensure that the test statistics $\wpstat$ and $\wjstat$ in Eq.~\eqref{eq:test_stat} are asymptotically equivalent to  
\bal\label{eq:kc_test_stat}
\kcwpstat 
&= n\E_{\vrobj'}\left(\int \w_{\vrobj'}(\vrad) 
\Diff_{\vrobj'}(\vrad) \prod_{k=1}^{\nsp}\diffop\mprf{k}_{\robj'_{k}}(\rad_{k})\right)\\
\text{and}\quad 
\kcwjstat 
&= n\E_{\vrobj'}\left(\int \w_{\vrobj'}(\vrad)
\Diff_{\vrobj'}(\vrad) \diffop\jprf_{\vrobj'}(\vrad)\right),\eal
respectively, 
where $\Diff_{\vrobj'}(\vrad) = \left\{
\sjprf_{\vrobj'}(\vrad) - \jprf_{\vrobj'}(\vrad) - \left[\prod_{k=1}^{\nsp}\smprf{k}_{\robj'_{k}}(\rad_k) - \prod_{k=1}^{\nsp}\mprf{k}_{\robj'_{k}}(\rad_k)\right]\right\}^2$,
and the limiting distribution of $\kcwpstat$ and $\kcwjstat$ can be obtained following Theorem~\ref{thm:sjprl}. 
Under Assumptions~\ref{ass:fcl_sjprf_brcktn_prime} and \ref{ass:weight}, we establish in Theorem~\ref{thm:null} that the proposed test statistics converge weakly to a mixture of countably many independent $\chi^2_{1}$ random variables, where the weights are determined by the weight profiles $\{\w_{\vobj}:\,\vobj\in\prodsp\}$ and the probabilities that $\robj_{k}$ falls in closed balls in $\msp_{k}$ and their pairwise intersections for $k=1,\dots,\nsp$. 

\begin{theorem}\label{thm:null}
Under $H_0$ in Eq.~\eqref{eq:hypo} and Assumptions~\ref{ass:fcl_sjprf_brcktn_prime} and \ref{ass:weight},  
the test statistics $\wpstat$ and $\wjstat$ in Eq.~\eqref{eq:test_stat} converge weakly to $\sum_{j=1}^{\infty} \wegnv_{j} \rnorm_{j}^2$ as $n\ra\infty$, 
where $\{\rnorm_{j}\}_{j=1}^{\infty}$ are independent standard normal random variables, 
and $\wegnv_{1}\ge \wegnv_{2}\ge \dots$ are the eigenvalues of $\covfwjstat\colon(\prodsp\times \qprfdom^{\nsp})^2\ra\real$ given by 
\bal\label{eq:gpcov}
&\covfwjstat(\vobj,\vu,\vobj',\vu')\\
&= \left(\w_{\vobj}\left\{ (\mprf{1}_{\obj_1})\inv(u_1),\dots, (\mprf{\nsp}_{\obj_{\nsp}})\inv(u_{\nsp}) \right\} \w_{\vobj'}\left\{ (\mprf{1}_{\obj'_1})\inv(u'_1),\dots, (\mprf{\nsp}_{\obj'_{\nsp}})\inv(u'_{\nsp}) \right\} \right)^{1/2}\\
&\quad\cdot\left\{\prod_{k=1}^{\nsp} \prob\left[d_k(\obj_k,\robj_k)\le(\mprf{k}_{\obj_k})\inv(u_k),\, d_k(\obj'_k,\robj_k)\le (\mprf{k}_{\obj'_k})\inv(u'_k)\right] \right.\\
&\quad + (\nsp-1)\prod_{k=1}^{\nsp}\left\{ \mprf{k}_{\obj_k}\left[(\mprf{k}_{\obj_k})\inv(u_k)\right] \mprf{k}_{\obj'_k}\left[(\mprf{k}_{\obj'_k})\inv(u'_k)\right]\right\}\\
&\quad -\left.\sum_{l=1}^{\nsp} \right(
\prob\left[d_l(\obj_l,\robj_l)\le (\mprf{l}_{\obj_l})\inv(u_l),\, d_l(\obj'_l,\robj_l)\le (\mprf{l}_{\obj'_l})\inv(u'_l)\right]\\
&\quad\cdot\left.\left.
\prod_{k\in[\nsp]\backslash\{l\}} \left\{ \mprf{k}_{\obj_k}\left[(\mprf{k}_{\obj_k})\inv(u_k)\right] \mprf{k}_{\obj'_k}\left[(\mprf{k}_{\obj'_k})\inv(u'_k)\right]\right\} \right)\right\},
\eal
for $\vobj,\vobj'\in\prodsp$ and $\vu,\vu'\in \qprfdom^{\nsp}$. 
\end{theorem}

Let $\limcdf(\cdot)$ be the cumulative distribution function (cdf) of $\sum_{j=1}^{\infty} \wegnv_{j} \rnorm_{j}^2$ as defined in Theorem~\ref{thm:null}. 
The critical value of the level $\alpha$ asymptotic test with statistics in Eq.~\eqref{eq:test_stat} is given by  
\bal\label{eq:cval}
\cval = \inf\{t\in\real:\, \limcdf(t)\ge 1-\alpha\}. 
\eal
At level $\alpha$, the asymptotic test using $\wpstat$ (or $\wjstat$) in Eq.~\eqref{eq:test_stat} rejects $H_0$ in Eq.~\eqref{eq:hypo} if $\wpstat>\cval$ (or $\wjstat>\cval$). 
Theorem~\ref{thm:null} provides the asymptotic guarantee of type I error control and implies that the proposed tests have asymptotically correct size under $H_0$. 

The weights $\{\wegnv_{j}\}_{j=1}^{\infty}$ appearing in the limiting distribution of the test statistics under $H_0$ in Theorem~\ref{thm:null} generally depend on the probability measures induced by $\robj_{k}$ for $k=1,\dots,\nsp$. 
However, in the special case where the trivial weight profiles are used, i.e., $\w_{\vobj}\equiv 1$ for all $\vobj\in\prodsp$, we establish in Corollary~\ref{cor:dfree} that the proposed tests become distribution-free provided that we further assume: 
\begin{assumption}\label{ass:prf_cont} 
The marginal distance profiles $\mprf{k}_{\obj_k}$ are continuous for all $\obj_k\in\msp_k$ and $k=1,\dots,\nsp.$ 
\end{assumption}
\begin{corollary}\label{cor:dfree}
With the trivial weight profiles $\w_{\vobj}\equiv 1$ for all $\vobj\in\prodsp$, under $H_0$ in Eq.~\eqref{eq:hypo} and the assumptions of Theorem~\ref{thm:null} as well as Assumption~\ref{ass:prf_cont}, 
the proposed tests with statistics $\wpstat$ and $\wjstat$ in Eq.~\eqref{eq:test_stat} are asymptotically distribution-free. 
Specifically, $\wpstat$ and $\wjstat$ converge weakly to $\sum_{j=1}^{\infty} \dfegnv_{j}\rnorm_{j}^2$, 
where $\{\rnorm_{j}\}_{j=1}^{\infty}$ are independent standard normal random variables, 
and $\dfegnv_{1}\ge \dfegnv_{2}\ge \dots$ are eigenvalues of $\dfcovfwjstat\colon \qprfdom^{\nsp}\times \qprfdom^{\nsp} \ra\real$ given by 
\bal\nn
\dfcovfwjstat(\vu,\vu')
&= \prod_{k\in[\nsp]}(u_k\wedge u'_k) - \prod_{k\in[\nsp]}(u_k u'_k) 
- \sum_{l=1}^{\nsp} 
\left\{ \left(u_l\wedge u'_l -u_l u'_l\right)
\prod_{k\in[\nsp]\backslash\{l\}} \left(u_k u'_k\right) \right\}
\eal
for $\vu,\vu'\in \qprfdom^{\nsp}$, with $u\wedge u' = \min\{u,u'\}$ for any $u,u'\in\real$.  
\end{corollary}

\begin{remark} \label{rem: bkr} 
Observe that for $\nsp=2$, with trivial weight profiles $\w_{\vobj}\equiv 1$ for all $\vobj\in\prodsp$, $\wjstat(\vrobj_1,\vrobj_2,\dots,\vrobj_n) 
= \sum_{i=1}^{n} \int  \{ \sjprf_{\vrobj_{i}}(\vrad) - \prod_{k=1}^{\nsp} \smprf{k}_{\robj_{i,k}}(\rad_{k}) \}^2 \diffop\sjprf_{\vrobj_{i}}(\vrad)$. 
Conditional on $\vrobj_{i}$, each summand $\int \{ \sjprf_{\vrobj_{i}}(\vrad) - \prod_{k=1}^{\nsp} \smprf{k}_{\robj_{i,k}}(\rad_{k}) \}^2 \diffop\sjprf_{\vrobj_{i}}(\vrad)$ 
corresponds to the Blum--Kiefer--Rosenblatt test statistic \citep{blum:61} 
for the independence of $d_1(\obj_{1},\robj_{1})$ and $d_2(\obj_{2},\robj_{2})$ at $\bm x =\vrobj_{i}$, 
which in turn is equivalent to Hoeffding's test of independence \citep{hoef:48}. 
This demonstrates that our proposed tests provide a natural extension of classical nonparametric measures of independence, 
originally developed for random vectors, to the more general setting of random objects. Although the classical results were established for the bivariate case ($\nsp = 2$), 
the equivalence extends seamlessly for $\nsp \geq 3$. Moreover, several desirable properties of the classical Blum--Kiefer--Rosenblatt statistic, including the distribution-free property of the asymptotic distribution, are preserved by $\wjstat$ with the trivial weight profiles. 
\end{remark}

In addition, we establish in Proposition~\ref{prop:wjstat_vs_Ewjstat} that $n\inv\wpstat$ and $n\inv\wjstat$ are consistent estimators of $\Ewpstat$ and $\Ewjstat$ in Eq.~\eqref{eq:Ewjstat}, respectively, where $\wpstat$ and $\wjstat$ are in Eq.~\eqref{eq:test_stat}. 
\begin{proposition}\label{prop:wjstat_vs_Ewjstat}
Under Assumptions~\ref{ass:fcl_sjprf_brcktn_prime} and \ref{ass:weight}, $n\inv\wpstat$ and $n\inv\wjstat$ converge in probability to $\Ewpstat$ and $\Ewjstat$, respectively, as $n\ra\infty$. 
\end{proposition}
Note that Proposition~\ref{prop:wjstat_vs_Ewjstat} does not require the assumption of mutual independence and serves as another pillar for the proposed tests for mutual independence using statistics $\wpstat$ and $\wjstat$ , which originate from the measures of mutual independence $\Ewpstat$ and $\Ewjstat$, respectively. 

Next, we establish the consistency of the proposed tests under the alternatives. 
We analyze the power of the proposed tests using $\wpstat$ and $\wjstat$ under sequences of local alternatives 
\bal\label{eq:alternative_seq}
\HaEwpstat 
= \left\{\prob\vrobj\inv:\, \Ewpstat = \Ewpstat_n\right\}
\quad\text{and}\quad
\HaEwjstat 
= \left\{\prob\vrobj\inv:\, \Ewjstat = \Ewjstat_n\right\}, 
\eal 
respectively, where $\Ewpstat_n$ and $\Ewjstat_n$ are sequences of positive numbers such that 
\bal\nn
\Ewpstat_n\ra 0,\quad 
n\Ewpstat_n\ra \infty, \quad\text{and}\quad 
\sup_{\vobj\in\jtsupp,\,\vrad\in\rdom} \left\lvert\sw_{\vobj}(\vrad) - \w_{\vobj}(\vrad)\right\rvert =\Op\left((\Ewpstat_n)^{1/2}\right), 
\eal 
and that 
\bal\nn
\Ewjstat_n\ra 0,\quad 
n\Ewjstat_n\ra \infty, 
\quad\text{and} \quad 
\sup_{\vobj\in\jtsupp,\,\vrad\in\rdom} \left\lvert\sw_{\vobj}(\vrad) - \w_{\vobj}(\vrad)\right\rvert =\Op\left((\Ewjstat_n)^{1/2}\right)
\eal 
respectively, as $n\ra\infty$. 
Hence, both $\HaEwpstat$ and $\HaEwjstat$ converge towards $H_0$ as $n\ra\infty$. 
The power of the level $\alpha$ asymptotic test using $\wpstat$ under $\HaEwpstat$ is given by $\prob_{\HaEwpstat}(\wpstat>\cval)$. 
Similarly, the power of the test using $\wjstat$ under $\HaEwjstat$ is given by $\prob_{\HaEwjstat}(\wjstat>\cval)$. 
We establish in Corollary~\ref{cor:power} that the proposed tests are consistent under the sequences of local alternatives shrinking towards $H_0$. 

\begin{corollary}\label{cor:power}
Under the assumptions of Theorem~\ref{thm:null}, $\prob_{\HaEwpstat}(\wpstat>\cval)\ra 1$ and $\prob_{\HaEwjstat}(\wjstat>\cval)\ra 1$ as $n\ra\infty$.
\end{corollary}

\begin{remark}\label{rmk:local_Ha}
We illustrate the local alternatives in Eq.~\eqref{eq:alternative_seq} by a bivariate Gaussian example. 
For any given $\rho\in(-1,1)$, consider $\vrobj=(\robj_1,\robj_2)\tps\sim N(\bm{0},\mathbf{\Sigma}_\rho)$ with $\mathbf{\Sigma}_\rho=\begin{pmatrix}1&\rho\\ \rho&1\end{pmatrix}$, and denote the corresponding $\Ewpstat$ and $\Ewjstat$ by $\Ewpstat_{\rho}$ and $\Ewjstat_{\rho}$, respectively. 
If the weight profiles $\w_{\vobj}$ are uniformly bounded from above and below by two positive constants over $\vobj\in\real^2$, i.e., there exist positive constants $\lcw$ and $\cw$ such that $\lcw\le \lvert\w_{\vobj}(\vrad)\rvert\le \cw$ for all $\vobj\in\real^2$ and $\vrad\in(0,+\infty)^2$, 
then by Lemma~S.2 
in the Supplement, there exists a constant $\kappa>0$ such that 
\bal\nn
\Ewpstat_{\rho} = \kappa\rho^2 + O(\rho^4)
\quad\text{and}\quad 
\Ewjstat_{\rho} = \kappa\rho^2 + O(\rho^4)
\quad \text{as } |\rho|\ra 0.
\eal
Hence, the local alternatives in Eq.~\eqref{eq:alternative_seq} are satisfied if and only if 
\bal\nn
|\rho|\ra 0, \quad n\rho^2 \ra\infty \quad\text{and}\quad \sup_{\vobj\in\jtsupp,\,\vrad\in\rdom} \left\lvert\sw_{\vobj}(\vrad) - \w_{\vobj}(\vrad)\right\rvert =\Op\left(|\rho|\right). 
\eal
Taking $\rho = \rho_n$ as any sequence that converges to $0$ more slowly than $n^{-1/2}$ as $n\ra\infty$ satisfies the local alternatives in Eq.~\eqref{eq:alternative_seq}; for instance, $\rho_n = n^{-1/2}\log(n)$ or $\rho_n = n^{-1/4}$. 
\end{remark}

While the proposed tests are shown to be consistent under the null and also the alternatives, obtaining the asymptotic critical value $\cval$ in Eq.~\eqref{eq:cval} is not straightforward in practice, in particular, when non-trivial weight profiles are considered. 
To this end, we propose to adopt a permutation scheme to approximate the critical value $\cval$ or to obtain $p$-values. 

Let $\sgrp$ be the set of all $n!$ permutations on the set $\{1,2,\dots,n\}$; specifically, each $\perm\in\sgrp$ is a bijective map from 
$\{1,2,\dots,n\}$ to itself. 
Let $\rperm$ be a random element following a uniform distribution on $\sgrp$ 
and $\rperm_{1,k},\rperm_{2,k},\dots,\rperm_{\nperm,k}$ for $k=2,3,\dots,\nsp$ be $\nperm(\nsp-1)$ independent copies of $\rperm$. 
Let $\rperm_{j,1}=\id$ be the identity map on $\{1,2,\dots,n\}$, for all $j=1,\dots,\nperm$. 
For each $j=1,\dots,\nperm$, applying the random permutation $\rperm_{j,k}$ to $\{\robj_{i,k}\}_{i=1}^{n}$ for each $k=1,\dots,\nsp$, respectively, yields a randomly permuted sample $\{\vrobj_{\rperm_{j,\cdot}(i)}\}_{i=1}^{n}$, where $\vrobj_{\rperm_{j,\cdot}(i)}=(\robj_{\rperm_{j,1}(i),1},\robj_{\rperm_{j,2}(i),2},\dots,\robj_{\rperm_{j,\nsp}(i),\nsp})\tps$; 
we denote by $\pwpstat{\rperm_{j,\cdot}}$ and $\pwjstat{\rperm_{j,\cdot}}$ the test statistics evaluated on $\{\vrobj_{\rperm_{j,\cdot}(i)}\}_{i=1}^{n}$ as per Eq.~\eqref{eq:test_stat}, i.e., 
\bal\nn
\pwpstat{\rperm_{j,\cdot}} 
&= \wpstat(\vrobj_{\rperm_{j,\cdot}(1)},\vrobj_{\rperm_{j,\cdot}(2)},\dots, \vrobj_{\rperm_{j,\cdot}(n)})\\
\text{and}\quad
\pwjstat{\rperm_{j,\cdot}} 
&= \wjstat(\vrobj_{\rperm_{j,\cdot}(1)},\vrobj_{\rperm_{j,\cdot}(2)},\dots,\vrobj_{\rperm_{j,\cdot}(n)}).
\eal  
Hence, the critical value $\cval$ in Eq.~\eqref{eq:cval} can be estimated by 
\bal\label{eq:pcval}
\pcvalwpstat = \inf\left\{t\in\real:\, \pcdfwpstat(t)\ge 1-\alpha\right\}
\ \text{and}\;\ 
\pcvalwjstat &= \inf\left\{t\in\real:\, \pcdfwjstat(t)\ge 1-\alpha\right\},
\eal
where 
\bal\label{eq:pcdf}
\pcdfwpstat(t) = \frac{1}{\nperm} \sum_{j=1}^{\nperm} \indicator\left\{\pwpstat{\rperm_j,\cdot} \le t\right\}\quad 
\text{and}\quad 
\pcdfwjstat(t) &= \frac{1}{\nperm} \sum_{j=1}^{\nperm} \indicator\left\{\pwjstat{\rperm_j,\cdot} \le t\right\}.
\eal
The $p$-values resulting from the permutation tests based on $\wpstat$ and $\wjstat$ in Eq.~\eqref{eq:test_stat} are respectively given by 
\bal\label{eq:ppval}
\ppvalwpstat &= \frac{1}{\nperm+1} \left( 1+\sum_{j=1}^{\nperm}\indicator\left\{\pwpstat{\rperm_j,\cdot} \ge \wpstat\right\} \right)\\
\text{and}\quad
\ppvalwjstat &= \frac{1}{\nperm+1} \left( 1+\sum_{j=1}^{\nperm}\indicator\left\{\pwjstat{\rperm_j,\cdot} \ge \wjstat\right\} \right).
\eal
We establish in Theorem~\ref{thm:perm_test} that the permutation tests with $p$-values in Eq.~\eqref{eq:ppval} can effectively control the type I error rate under $H_0$ in Eq.~\eqref{eq:hypo} and also are consistent under the sequences of local alternatives $\HaEwpstat$ and $\HaEwjstat$ in Eq.~\eqref{eq:alternative_seq}. 

\begin{theorem}\label{thm:perm_test}
Under the assumptions of Theorem~\ref{thm:null}, 
\begin{enumerate}[(\roman*)]
    \item $\lim_{n\ra\infty,\,\nperm\ra\infty}\prob_{H_0}(\ppvalwpstat\le\alpha)\le \alpha$ and $\lim_{n\ra\infty,\,\nperm\ra\infty}\prob_{H_0}(\ppvalwjstat\le\alpha)\le \alpha$; 
    \item $\prob_{\HaEwpstat}(\ppvalwpstat\le \alpha)\ra 1$ and $\prob_{\HaEwjstat}(\ppvalwjstat\le \alpha)\ra 1$ as $n\ra\infty$. 
\end{enumerate}
\end{theorem}

In addition, we establish that the proposed tests achieve the minimax rate optimality. 
For any given $\eps>0$, consider the alternative hypotheses $\hap{\eps}:\, \Ewpstat > \eps$ and $\haj{\eps}:\, \Ewjstat > \eps$. 
Let $\testclass(\alpha)$ be the class of all level $\alpha$ test functions $\decision\colon (\prodsp)^n\ra\{0,1\}$, which maps the input data $\{\vrobj_i\}_{i=1}^{n}$ to $0$ for failing to reject $H_0$ or $1$ for rejecting $H_0$. 
The minimax type II error rates for the alternatives in $\hap{\eps}$ and $\haj{\eps}$ are defined as 
\bal\nn
\tIIerp(\eps) 
&= \inf_{\decision\in\testclass(\alpha)} \sup_{\prob\vrobj\inv\in\hap{\eps}} \prob\left\{\decision(\vrobj_1,\dots,\vrobj_n) = 0\right\}
\\ \text{and}\quad 
\tIIerj(\eps) 
&= \inf_{\decision\in\testclass(\alpha)} \sup_{\prob\vrobj\inv\in\haj{\eps}} \prob\left\{\decision(\vrobj_1,\dots,\vrobj_n) = 0\right\},
\eal
respectively. 
We establish in Theorem~\ref{thm:minimax} that if $\eps$ is not larger in order than $n\inv$, the minimax type II error rates for all level $\alpha$ tests are bounded away from $0$. 
\begin{theorem}\label{thm:minimax}
If the weight profile $\w$ is uniformly away from $0$, i.e., $\w_{\vobj}(\vrad)\ge \lcw$ for all $\vobj\in\supp(\prob\vrobj\inv)$ and $\vrad\in\rdom$ for some $\lcw > 0$, then 
for any $\zeta\in(0,1-\alpha)$, there exists a constant $c_{\alpha,\zeta}>0$ such that $\tIIerp(cn\inv)\ge \zeta$ and $\tIIerj(cn\inv)\ge \zeta$ for all $c\in(0,c_{\alpha,\zeta})$. 
\end{theorem}
The results in Corollary~\ref{cor:power} and Theorem~\ref{thm:perm_test} in conjunction with Theorem~\ref{thm:minimax} imply that the proposed tests attain minimax rate optimality.

\section{Simulations}\label{sec:simulations}

In this section, we study the empirical performance of our proposed tests across a range of diverse scenarios, including both Euclidean data and non-Euclidean random objects such as distributional data and random networks. 
In the family of the proposed tests, we consider $\wjstat$ with the weight profiles given by $\sw_{\vrobj_{i}}(\vrad) \equiv 1$, $\sw_{\vrobj_{i}}(\vrad) = \prod_{k=1}^{\nsp} \{ \smprf{k}_{\robj_{i,k}}(\rad_{k}) (1-\smprf{k}_{\robj_{i,k}}(\rad_{k})) \}^{-1}$ and $\sw_{\vrobj_{i}}(\vrad) = \prod_{k=1}^{\nsp} \{ \smprf{k}_{\robj_{i,k}}(\rad_{k}) \}^{-1}$ respectively, referred to as ``\jt'', ``\jtAD'', 
and ``\jtF'', respectively, 
as well as the counterparts of $\wpstat$ with the corresponding weight profiles,  referred to as ``\product'', ``\productAD'', and ``\productF'', respectively. We adopt the convention that $0/0=0$. 
In order to accommodate the data-dependent weight profiles, we implement all versions of the proposed tests using the permutation $p$-values defined in Eq.~\eqref{eq:ppval} with $B=500$ permutations. 
The code for implementing the proposed tests are available at \url{https://github.com/paromita1991/DiPMInd-code}. 
For $p=2$, we compare the power of our tests with the distance covariance \citep{szek:07}, denoted as ``\dcov'', implemented using the R package \texttt{energy} \citep{ener:22}, and the d-variable Hilbert Schmidt independence criterion \citep{pfis:18}, denoted as ``\dhsic'', implemented using the R package \texttt{dHSIC} \citep{dhsi:19}. For the latter, we employ the Laplacian kernel with the widely used median-of-distances heuristic for bandwidth selection, as recommended in \cite{pfis:18}. 
We also conduct mutual independence tests for $p=3$, where we compare the power of our tests not only with \dhsic, but also with the generalized distance covariance \citep{jin:18}, referred to as ``\mdcov'' hereafter, which extends distance covariance to handle mutual independence among more than two random vectors. 
We adopt the $l_2$ metric for random vectors in the Euclidean examples unless otherwise noted, specify the metric choices for each non-Euclidean example, and use the same metric for all the tests considered. 
We set the sample size $n=100$. 
We estimate the power of the tests using the proportion of rejections at a significance level of $0.05$ across $200$ Monte Carlo runs.

We introduce some notations to describe the examples.
Let $\gc{\mathbf{\Sigma}}{(F_1,\dots,F_l)}$ denote an $l$-dimensional Gaussian copula \citep{nels:06} with given marginal continuous cdfs $F_1, \dots, F_l$. 
To generate a random vector $\rvecbm \in \mathbb{R}^l$ such that $\rvecbm \sim \gc{\mathbf{\Sigma}}{(F_1,\dots,F_l)}$, we first generate $\bm{Z} = ({Z}_1, \dots, {Z}_l)\tps \sim N(\mathbf{0},\mathbf{\Sigma})$, and set $\rvecbm = (F_1^{-1}\{\Phi({Z}_1)\}, \dots, F_l^{-1}\{\Phi({Z}_l)\})\tps$, where $\Phi(\cdot)$ is the standard normal cdf and, for $j=1,\dots,l$, $F_j^{-1}(\cdot)$ is the left-continuous inverse of $F_j(\cdot)$ given by $F_j^{-1}(u) = \inf \{x \in \mathbb{R} : F_j(x) \geq u\}$ for $u \in (0,1)$. Here, $\mathbf{\Sigma}$ is an $l \times l$ covariance matrix controlling the dependency structure in the Gaussian copula. 
To explore various dependence structures, we will design simulations using covariance matrices $\mathbf{\Sigma}(\rho)_{m,l}=\begin{pmatrix}
(1 - \rho) \mathbf{I}_{m} + \rho \bm{1}_{m}\bm{1}_{m}\tps & \mathbf{O}_{m\times(l-m)} \\
\mathbf{O}_{(l-m)\times m} & \mathbf{I}_{l-m}
\end{pmatrix}
$ and $\mathbf{\Sigma}(\rho)_{l}=
(1 - \rho) \mathbf{I}_{l} + \rho \bm{1}_{l}\bm{1}_{l}\tps$. 
We will begin by exploring non-Euclidean settings followed by investigating the Euclidean settings for $p = 2$.

\begin{figure}[!hbt]
\centering
\includegraphics[width=\textwidth]{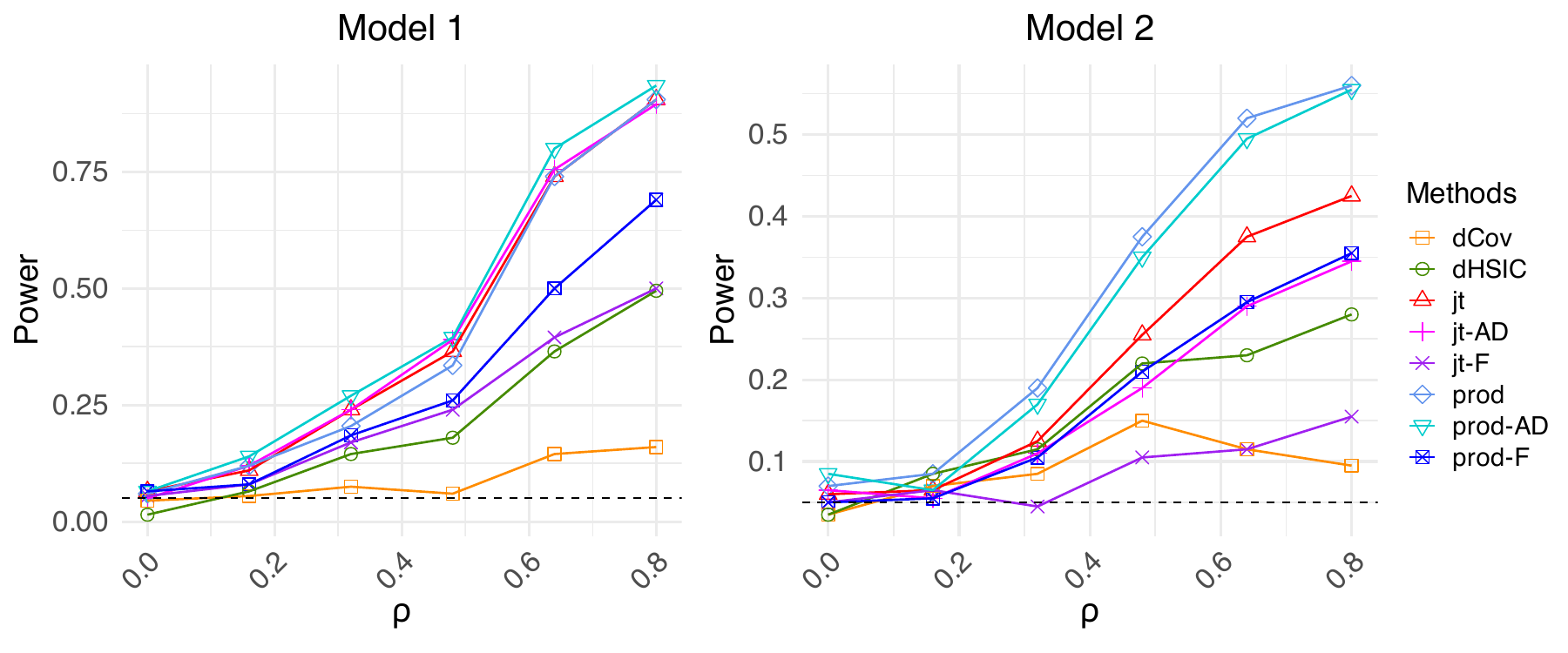}
 \caption{Empirical power as a function of $\rho$ of the different tests in the setting of Example~\ref{ex:prefattachment} for random networks ($\robj_{i,1}$) paired with random vectors in $\mathbb{R}^{l}$ ($\robj_{i,2}$). In Model~1, $l=40$ and $\robj_{i,2}=(\rvec_{i,11}, \dots, \rvec_{i,50})\tps$ with $\rvecbm_i \sim N(\mathbf{0}_{50}, \mathbf{\Sigma}(\rho)_{20,50})$ while $l=30$ in Model~2, and $\robj_{i,2}=(\rvec_{i,21}, \dots, \rvec_{i,50})\tps$ with $\rvecbm_i \sim N(\mathbf{0}_{50}, \mathbf{\Sigma}(\rho)_{40,50})$. Networks with 200 nodes, represented as graph Laplacian matrices, are generated using the preferential attachment model \citep{bara:99}, with attachment proportional to $(\mathrm{vertex\ degree})^{\gamma_i}$. For Model~1, $\gamma_i= (\{1+\exp(-\rvec_{i,1}\rvec_{i,11}) \}^{-1}, \dots,  \{1+\exp(-\rvec_{i,10}\rvec_{i,20}) \}^{-1})\tps Y$ and $Y=(U_1, \dots, U_{10})\tps/(U_1+\dots+U_{10})$ where $U_1, \dots, U_{10}$ are generated independently from $\mathrm{Unif}(0,1)$. For Model~2, $\gamma_i= \operatorname{median} \{\{1+\exp(-W_i \rvec_{i,1})\}^{-1}, \dots, \{1+\exp(-W_i \rvec_{i,20})\}^{-1}\}$ with $W_i=(2\mathbb{I}(\rvec_{i,21}<1)-1)$.  The black dashed line indicates the significance level $\alpha = 0.05$.}
 \label{fig:prefattachment}
\end{figure}

\begin{example}\label{ex:prefattachment}
    For $i =1, \dots, n$, we generate $\vrobj_{i} = (\robj_{i,1},\robj_{i,2})\tps \in \msp_1 \times  \mathbb{R}^{l}$ where $\msp_1$ is the space of networks with 200 nodes represented as graph Laplacian matrices. In this example, we use the $l_{1}$ metric for $\mathbb{R}^{l}$, and for graph Laplacian matrices $\mathbf{L} = (L_{j,j'})$ and $\wt{\mathbf{L}}= (\wt{L}_{j,j'})$, we adopt the metric given by $d_1(\mathbf{L},\wt{\mathbf{L}})=\sum_{j,j'=1}^{200} \lvert L_{j,j'}-\wt{L}_{j,j'} \rvert$. Two models are considered. In Model 1, we set $l=40$. We generate $\rvecbm_i \sim N(\mathbf{0}_{50}, \mathbf{\Sigma}(\rho)_{20,50})$ and let $\robj_{i,2}=( \rvec_{i,11}, \dots,  \rvec_{i,50})\tps$. The random networks $\robj_{i,1}$ are generated according to the preferential attachment model \citep{bara:99} where the attachment function is proportional to the vertex degree to the power of $\gamma_i$. 
    We generate $\gamma_i= (\{1+\exp(-\rvec_{i,1}\rvec_{i,11}) \}^{-1}, \dots,  \{1+\exp(-\rvec_{i,10}\rvec_{i,20}) \}^{-1})\tps Y$, where $Y=(U_1, \dots, U_{10})\tps/(U_1+\dots+U_{10})$ with $U_1, \dots, U_{10}$ generated independently from $\mathrm{Unif}(0,1)$. 
    In Model 2, we set $l=30$ and generate $\robj_{i,2}=( \rvec_{i,21}, \dots,  \rvec_{i,50})\tps$ with $\rvecbm_i \sim N(\mathbf{0}_{50}, \mathbf{\Sigma}(\rho)_{40,50})$. 
    We generate $\gamma_i$ as $\gamma_i= \operatorname{median} \{\{1+\exp(-W_i \rvec_{i,1})\}^{-1}, \dots, \{1+\exp(-W_i \rvec_{i,20})\}^{-1}\}$ with $W_i=(2\mathbb{I}(\rvec_{i,21}<1)-1)$.
    For generating the random networks, we used the function \texttt{sample\_pa()} in the R package \texttt{igraph} (version 1.3.5) \citep{igra:06}. We present the empirical power curves of the different tests as a function of $\rho$ in Figure~\ref{fig:prefattachment}. 
    The proposed tests outperform the competitors for both models. In particular, two new tests among the proposed tests, \product and \productAD, demonstrate the best performance in Model~2.
\end{example}

\begin{figure}[!hbt]
 \centering
 \includegraphics[width=\textwidth]{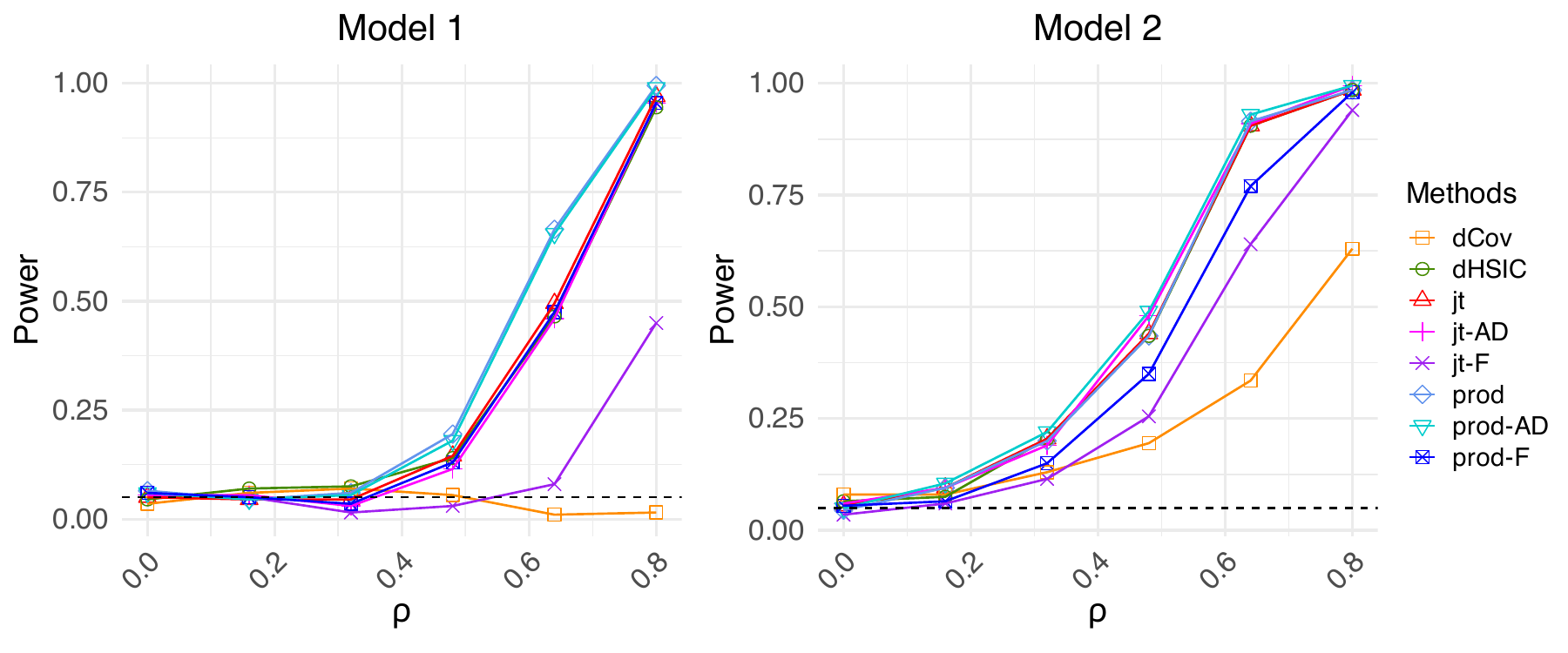}
 \caption{Empirical power as a function of $\rho$ for testing independence between objects in $\msp_1$, the space of networks with 60 nodes, and $\msp_2$, the space of networks with 100 nodes, generated according to stochastic block models \citep{holl:83} as described in Example~\ref{ex:sbm}. Networks are parameterized by connectivity matrices $\mathbf{A}_{1,i}$ (three blocks: $30$, $15$, $15$ nodes) and $\mathbf{A}_{2,i}$ (five blocks: $20$ nodes each) that depend on $\rvecbm_i \in \mathbb{R}^{12}$, with $\mathbf{A}_{1,i} = \diag\left(\frac{0.75}{1 + e^{-(\rvec_{i,1} \rvec_{i,2})^{-1}}}, \frac{0.75}{1 + e^{-(\rvec_{i,3} \rvec_{i,4})^{-1}}}, \frac{0.75}{1 + e^{- (\rvec_{i,5} \rvec_{i,6}\rvec_{i,7})^{-1} }}\right) + 0.1 \mathbf{1}_{3}\mathbf{1}_{3}\tps$ and $\mathbf{A}_{2,i} = \diag\left(\frac{0.75}{1 + e^{-\rvec_{i,8}}}, \dots, \frac{0.75}{1 + e^{-\rvec_{i,12}}}\right) + 0.05 \mathbf{1}_{5}\mathbf{1}_{5}\tps$ (Model~1) in the left panel and $\mathbf{A}_{1,i}=\diag\left(
\frac{0.75}{1 + e^{- \rvec_{i,1} \rvec_{i,2}}},
\frac{0.75}{1 + e^{- \rvec_{i,3} \rvec_{i,4}}},
\frac{0.75}{1 + e^{-\lvert \rvec_{i,5}\rvec_{i,6} \rvec_{i,7} \rvert}}
\right) + 0.1 \mathbf{1}_{3}\mathbf{1}_{3}\tps$ and $\mathbf{A}_{2,i}$ same as above (Model~2) in the right panel. The vectors $\rvecbm_i \sim \gc{\mathbf{\Sigma}(\rho)_{10,12}}{(F_1, \dots, F_{12})}$, with $F_1, \dots, F_{10}$ as cdf of $t$-distributions with 2 degrees of freedom and $F_{11}, F_{12}$ as the cdf of $N(0,1)$. The black dashed line indicates the significance level $\alpha = 0.05$.}
 \label{fig:sbm}
\end{figure}

\begin{example}\label{ex:sbm}
    For $i =1, \dots, n$, we generate $\vrobj_{i} = (\robj_{i,1},\robj_{i,2})\tps \in \msp_1 \times  \msp_2 $ where $\msp_1$ and $\msp_2$ are the spaces of graph Laplacian matrices of networks with $k_1=60$ and $k_2=100$ nodes, respectively. We adopt the metric $d_k(\mathbf{L},\wt{\mathbf{L}}) = \sum_{j,j'=1}^{k} \lvert L_{j,j'}-\wt{L}_{j,j'} \rvert$ for $k\times k$ graph Laplacian matrices $\mathbf{L}$ and $\wt{\mathbf{L}}$. 
    First we generate $\rvecbm_i \in \mathbb{R}^{12}$ such that $\rvecbm_i \sim \gc{\mathbf{\Sigma}(\rho)_{10,12}}{(F_1,\dots,F_{12})}$, with $F_j$ being the cdf of the $t$-distribution with two degrees of freedom for $j=1, \dots, 10$ and the cdf of $N(0,1)$ for $l=11,12$. 
    We consider two models. In Model~1, we generate the random networks $\robj_{i,1}$ (resp. $\robj_{i,2}$) according to the stochastic block model \citep{holl:83} with three blocks of sizes $30,15$ and $15$ nodes each (resp. five blocks each of size $20$ nodes). 
    Here, the block connectivity matrices $\mathbf{A}_{1,i}$ and $\mathbf{A}_{2,i}$ for $\robj_{i,1}$ and $\robj_{i,2}$, respectively, are given by $\mathbf{A}_{1,i} = \diag\left(\frac{0.75}{1 + e^{-(\rvec_{i,1} \rvec_{i,2})^{-1}}}, \frac{0.75}{1 + e^{-(\rvec_{i,3} \rvec_{i,4})^{-1}}}, \frac{0.75}{1 + e^{- (\rvec_{i,5} \rvec_{i,6}\rvec_{i,7})^{-1} }}\right) + 0.1 \mathbf{1}_{3}\mathbf{1}_{3}\tps$ and $\mathbf{A}_{2,i}=\diag\left(\frac{0.75}{1 + e^{-\rvec_{i,8}}},
    \frac{0.75}{1 + e^{-\rvec_{i,9}}},
\frac{0.75}{1 + e^{-\rvec_{i,10}}},
\frac{0.75}{1 + e^{-\rvec_{i,11}}},
\frac{0.75}{1 + e^{-\rvec_{i,12}}}
\right) + 0.05 \mathbf{1}_{5}\mathbf{1}_{5}\tps$. 
We present the empirical power curves of the different tests as a function of $\rho$ in the left panel of Figure~\ref{fig:sbm}. 
In the right panel we illustrate the results for Model~2, where $\robj_{i,1}$  is generated with $\mathbf{A}_{1,i}=\diag\left(
\frac{0.75}{1 + e^{- \rvec_{i,1} \rvec_{i,2}}},
\frac{0.75}{1 + e^{- \rvec_{i,3} \rvec_{i,4}}},
\frac{0.75}{1 + e^{-\lvert \rvec_{i,5}\rvec_{i,6} \rvec_{i,7} \rvert}}
\right) + 0.1 \mathbf{1}_{3}\mathbf{1}_{3}\tps$ and everything else is kept in the same way as Model~1. 
For simulating the random networks we used the function \texttt{sampleSimpleSBM} in the R package \texttt{sbm} (version 0.4.7) \citep{sbm:24}. 
The proposed tests have outstanding performance for both models. In particular, the \product and \productAD versions of the proposed tests demonstrate the best performance in Model 1, outperforming all other tests.
\end{example}
  
\begin{figure}[!hbt]
\centering
\includegraphics[width=\textwidth]{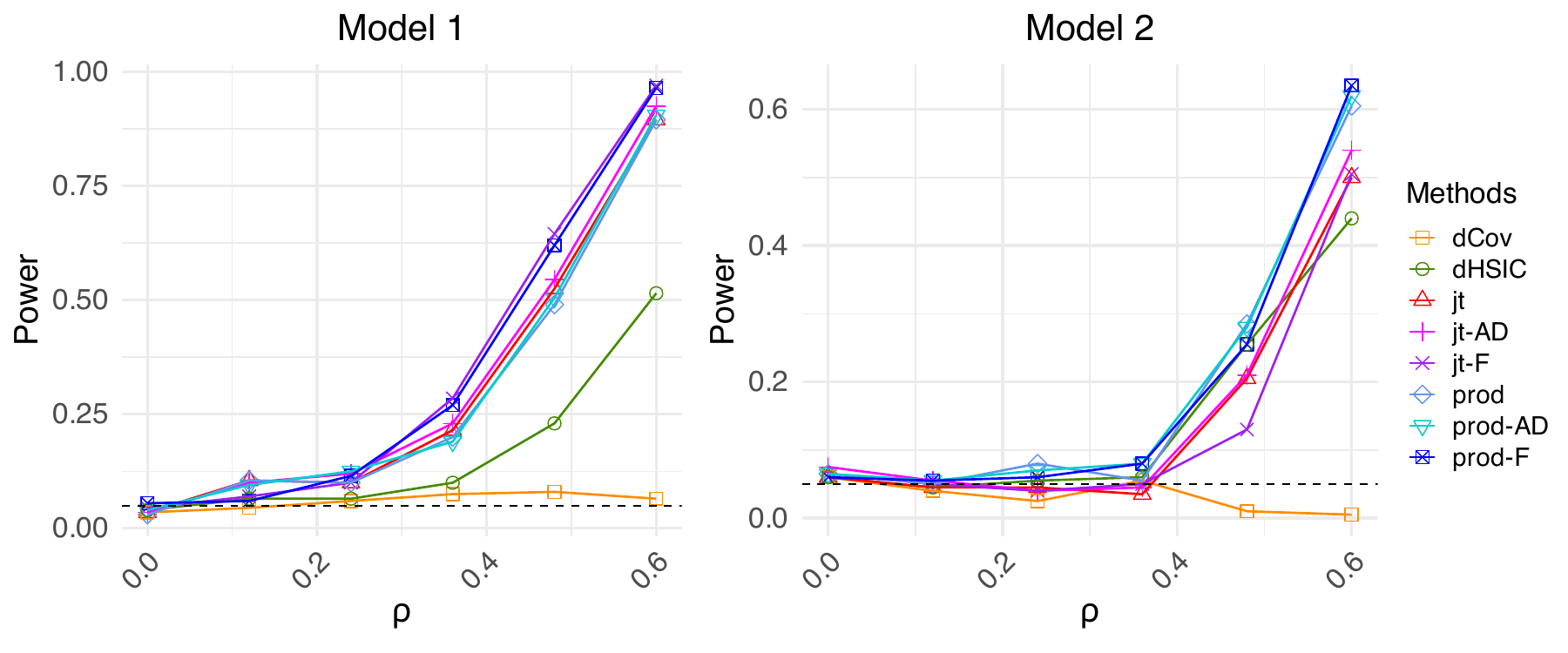}
 \caption{Empirical power as a function of $\rho$ for the different tests involving $\vrobj_i = (\robj_{i,1}, \robj_{i,2})\tps \in \msp_1 \times \mathbb{R}^{14}$, where $\msp_1$ is the space of bivariate normal distributions as described in Example~\ref{ex:distributionaldata}. Vectors $\robj_{i,2}=( \rvec_{i,7}^3, \dots,  \rvec_{i,20}^3)\tps$ are derived from $\rvecbm_i \sim \gc{\mathbf{\Sigma}(\rho)_{20}}{(F, \dots, F)}$, with $F$ being the cdf of a $t$-distribution with 2 degrees of freedom. 
 For each $i=1,\dots,n$, $\robj_{i,1}=N(\bm\mu_i, \mathbf{\Sigma}_i)$, where $\bm\mu_i=\left(
\rvec_{i,1} (1+\rvec_{i,5})^{-1} + 0.1 \epsilon_{1,i},
\rvec_{i,2} (1+\rvec_{i,6})^{-1} + 0.1 \epsilon_{2,i}
\right)\tps$ in Model 1 shown in the left panel and $\bm\mu_i=\left(
\rvec_{i,1}^{-1} (1+\rvec_{i,5}^{-1})^{-1} + 0.1 \epsilon_{1,i},
\rvec_{i,2}^{-1} (1+\rvec_{i,6}^{-1})^{-1} + 0.1 \epsilon_{2,i}
\right)\tps$ in Model 2 shown in the right panel, 
 $\epsilon_{1,i}, \epsilon_{2,i} \sim N(0,1)$ independently, and $\mathbf{\Sigma}_i = \diag\left(\{1 + e^{-\rvec_{i,3}^2}\}\inv, \{1 + e^{-\rvec_{i,4}^2}\}\inv\right)$ in both Model 1 and Model 2. The black dashed line indicates the significance level $\alpha = 0.05$. }
 \label{fig:distributionaldata}
\end{figure}

\begin{example}\label{ex:distributionaldata}
For $i =1, \dots, n$, we generate $\vrobj_{i} = (\robj_{i,1},\robj_{i,2})\tps \in \msp_1 \times  \mathbb{R}^{14} $ where $\msp_1$ is the space of bivariate normal distributions, endowed with the 2-Wasserstein metric given by $W_2(N(\bm\mu_1, \mathbf{\Sigma}_1),N(\bm\mu_2, \mathbf{\Sigma}_2))=
\{\| \bm\mu_1 - \bm\mu_2 \|^2 + \| \mathbf{\Sigma}_1^{1/2} - \mathbf{\Sigma}_2^{1/2} \|_F^2\}^{1/2}
$ with $\| \mathbf{A} \|_F$ being the Frobenius norm of any matrix $\mathbf{A}$.  
First we generate $\rvecbm_i \in \mathbb{R}^{20}$ such that $\rvecbm_i \sim \gc{\mathbf{\Sigma}(\rho)_{20}}{(F,\dots,F)}$, with $F$ being the cdf of the $t$-distribution with two degrees of freedom and set $\robj_{i,2}=( \rvec_{i,7}^3, \dots,  \rvec_{i,20}^3)\tps$. Next we generate $\robj_{i,1}$ to be a random bivariate normal distribution $N(\bm\mu_i, \mathbf{\Sigma}_i)$ such that $\bm\mu_i=\left(
\rvec_{i,1} (1+\rvec_{i,5})^{-1} + 0.1 \epsilon_{1,i},
\rvec_{i,2} (1+\rvec_{i,6})^{-1} \right.\newline+\left. 0.1 \epsilon_{2,i}
\right)\tps$ and $\mathbf{\Sigma}_i=\diag\left(
\frac{1}{1 + e^{-(\rvec_{i,3})^2}},
\frac{1}{1 + e^{-(\rvec_{i,4})^2}}
\right)$ with $\epsilon_{1,i}$ and $\epsilon_{2,i}$ independently generated from $N(0,1)$ (Model~1). We present the empirical power curves of the different tests as a function of $\rho$ in the left panel of Figure~\ref{fig:distributionaldata}. In the right panel, we investigate Model~2 where we generate $\bm\mu_i=\left(
\rvec_{i,1}^{-1} (1+\rvec_{i,5}^{-1})^{-1} + 0.1 \epsilon_{1,i},
\rvec_{i,2}^{-1} (1+\rvec_{i,6}^{-1})^{-1} + 0.1 \epsilon_{2,i}
\right)\tps$ while keeping everything else the same as in Model~1. 
The proposed tests outperform all the competitors in Model~1 and the \productAD and \productF tests dominate the power performance in Model~2.
\end{example}

\begin{figure}[!hbt]
 \centering
\includegraphics[width=0.55\textwidth]{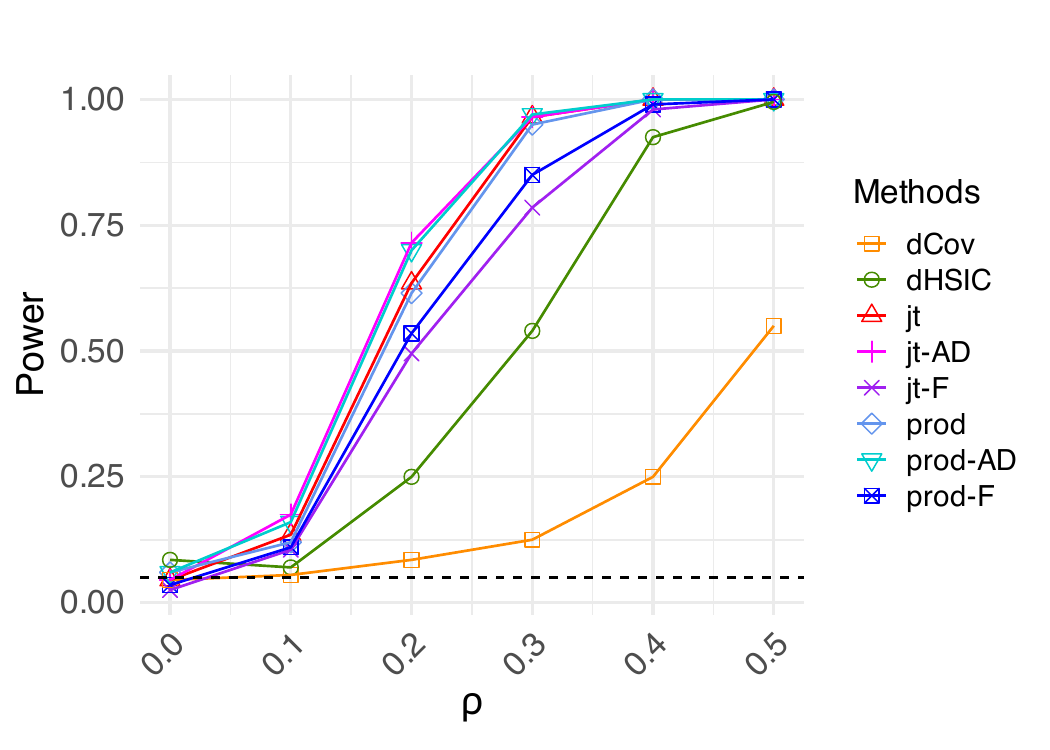}
 \caption{Empirical power as a function of $\rho$ for the different tests involving $\vrobj_i = (\robj_{i,1}, \robj_{i,2})\tps \in \msp_1 \times \mathbb{R}^{35}$, where $\msp_1$ is the space of compositional vectors in $\mathbb{R}^{15}$ with non-negative entries summing to one. Data is generated as per Example~\ref{ex:compositionaldata} where vectors $\robj_{i,2} = (\rvec_{i,16}, \dots, \rvec_{i,50})\tps$ are derived from $\rvecbm_i \sim \gc{\mathbf{\Sigma}(\rho)_{50}}{(F, \dots, F)}$, with $F$ as the cdf of a $t$-distribution with three degrees of freedom. Compositional vectors $\robj_{i,1}$ are generated as $\robj_{i,1} = \operatorname{softmax}(\rvec_{i,1}^2, \dots, \rvec_{i,15}^2)\tps$, where $\operatorname{softmax}(\bm{z}) = \left(\frac{e^{z_1}}{\sum_{j=1}^l e^{z_j}}, \dots, \frac{e^{z_l}}{\sum_{j=1}^l e^{z_j}}\right)\tps$ for $\bm{z}=(z_1, \dots, z_l)\tps$. The black dashed line indicates the significance level $\alpha = 0.05$.
}
\label{fig:compositionaldata}
\end{figure}

\begin{example}\label{ex:compositionaldata}
    For $i =1, \dots, n$, we generate $\vrobj_{i} = (\robj_{i,1},\robj_{i,2})\tps \in \msp_1 \times \mathbb{R}^{35} $ where $\msp_1$ is the space of compositional vectors in $\mathbb{R}^{15}$, that is, random vectors with non-negative entries that sum to one. We adopt the $l_1$ metric for $\mathbb{R}^{35}$. For any two compositional vectors $\bm{y}_1$ and $\bm{y}_2$ in $\msp_1$ we use the metric $d_c(\bm{y}_1,\bm{y}_2)=\arccos{(\sqrt{\bm{y}_1}\tps\sqrt{\bm{y}_2})}$.
    First we simulate $\rvecbm_i \in \mathbb{R}^{50}$ such that $\rvecbm_i \sim \gc{\mathbf{\Sigma}(\rho)_{50}}{(F,\dots,F)}$, with $F$ being the cdf of the $t$-distribution with three degrees of freedom and set $\robj_{i,2}=( \rvec_{i,16}, \dots,  \rvec_{i,50})\tps$. Then we generate $\robj_{i,1}$ as $\robj_{i,1}=\operatorname{softmax}\left(\rvec_{i,1}^2, \rvec_{i,2}^2, \ldots, \rvec_{i,15}^2\right)\tps$ where $\operatorname{softmax}(\bm{z})= \left(\frac{e^{z_1}}{\sum_{j=1}^l e^{z_j}}, \dots, \frac{e^{z_l}}{\sum_{j=1}^l e^{z_j}}\right)\tps$ for $\bm{z}=(z_1, \dots, z_l)\tps$. We present the empirical power curves of the different tests as a function of $\rho$ in Figure~\ref{fig:compositionaldata}. The proposed tests perform better than the competitors in this example.
\end{example}

Examples~\ref{ex:prefattachment}--\ref{ex:compositionaldata} illustrate that the proposed tests are powerful to detect independence in several non-Euclidean scenarios. 
We also explore examples in Euclidean scenarios, with Examples~\ref{ex:outlier}--\ref{ex:heavytailed} listed below and Examples~S.1--S.5 
deferred to Section~S.2 
in the Supplement. 

\begin{figure}[!hbt]
 \centering
\includegraphics[width=\textwidth]{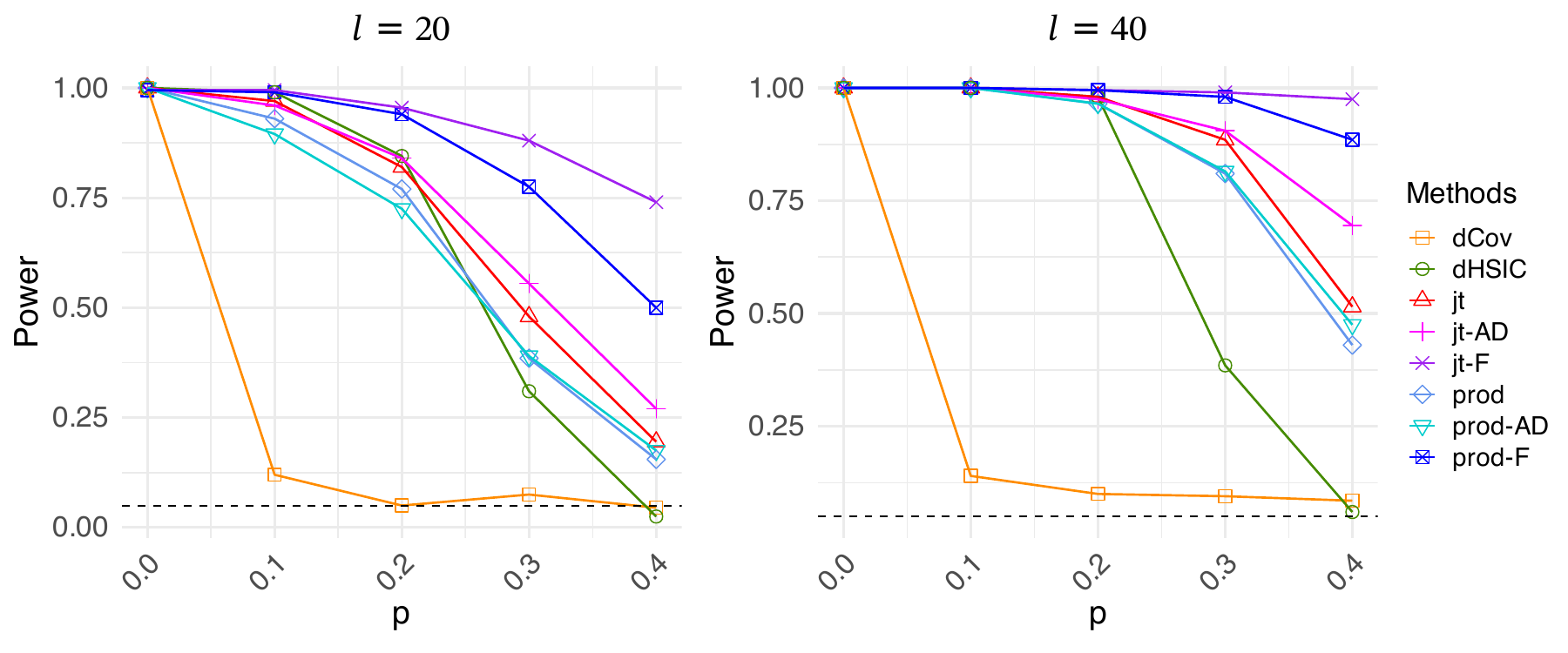}
 \caption{Empirical power as a function of the proportion $p$ for dimensions $l = 20$ and $l = 40$ in the scenario of Example~\ref{ex:outlier}. Data is generated as $\vrobj_i = (\robj_{i,1}, \robj_{i,2})\tps \in \mathbb{R}^{0.1l} \times \mathbb{R}^{0.9l}$, where $\rvecbm_i \in \mathbb{R}^l$ is sampled such that $\rvecbm_i \sim \gc{\mathbf{\Sigma}(0.6)_{0.2l,l}}{(F, \dots, F)}$, with $F$ being the cdf of $N(0,1)$. The components are defined as $\robj_{i,1} = (\rvec_{i,1}, \dots, \rvec_{i,0.1l})\tps$ and $\robj_{i,2} = (\rvec_{i,0.1l+1}, \dots, \rvec_{i,l})\tps + 200 \mathbb{I}\{U_i < p \} \bm{B}_i \odot \bm\epsilon_i$, where $U_i \sim \mathrm{Unif}(0,1)$, and $\bm{B}_i$ and $\bm\epsilon_i$ are random vectors in $\mathbb{R}^{0.9l}$ with components independently generated as $\mathrm{Bernoulli}(0.2)$ and $N(1,0.5^2)$ respectively, and $\bm{B}_i \odot \bm\epsilon_i$ denotes the element-wise product of $\bm{B}_i$ and $\bm\epsilon_i$. The black dashed line indicates the significance level $\alpha = 0.05$.}
 \label{fig:outlier}
\end{figure}

\begin{example}\label{ex:outlier}
    For $i =1, \dots, n$, we generate $\vrobj_{i} = (\robj_{i,1},\robj_{i,2})\tps \in \mathbb{R}^{0.1l} \times \mathbb{R}^{0.9l}$ by first obtaining $\rvecbm_i \in \mathbb{R}^l$ such that $\rvecbm_i \sim \gc{\mathbf{\Sigma}(0.6)_{0.2l,l}}{(F,\dots,F)}$, with $F$ being the cdf of the $N(0,1)$, and then we set $\robj_{i,1}=( \rvec_{i,1}, \dots,  \rvec_{i,0.1l})\tps$, and $\robj_{i,2}=( \rvec_{i,0.1l+1}, \dots,  \rvec_{i,l})\tps + 200 \mathbb{I}\{U_i < p \} \bm{B}_i \odot \bm\epsilon_i $ 
    Here $U_i \sim \mathrm{Unif}(0,1)$, and $\bm{B}_i$ and $\bm\epsilon_{i}$ are random vectors in $\mathbb{R}^{0.9l}$ whose components are independently generated as $\mathrm{Bernoulli}(0.2)$ and $N(1,0.5^2)$ respectively, $\bm{B}_i \odot \bm\epsilon_i$ denotes the element-wise product of $\bm{B}_i$ and $\bm\epsilon_i$. We adopt the $l_1$ metric between the random vectors to evaluate the different tests in this example. In Figure~\ref{fig:outlier}, we present the empirical power curves as a function of the proportion $p$ when $l=20$ and $l=40$. The proposed tests, in particular the \jtF and \productF versions, are robust to data corruption by outliers and demonstrate excellent performance even with large proportion of outliers as opposed to the competing tests.
\end{example}

\begin{figure}[!hbt]
 \centering
 \includegraphics[width=\textwidth]{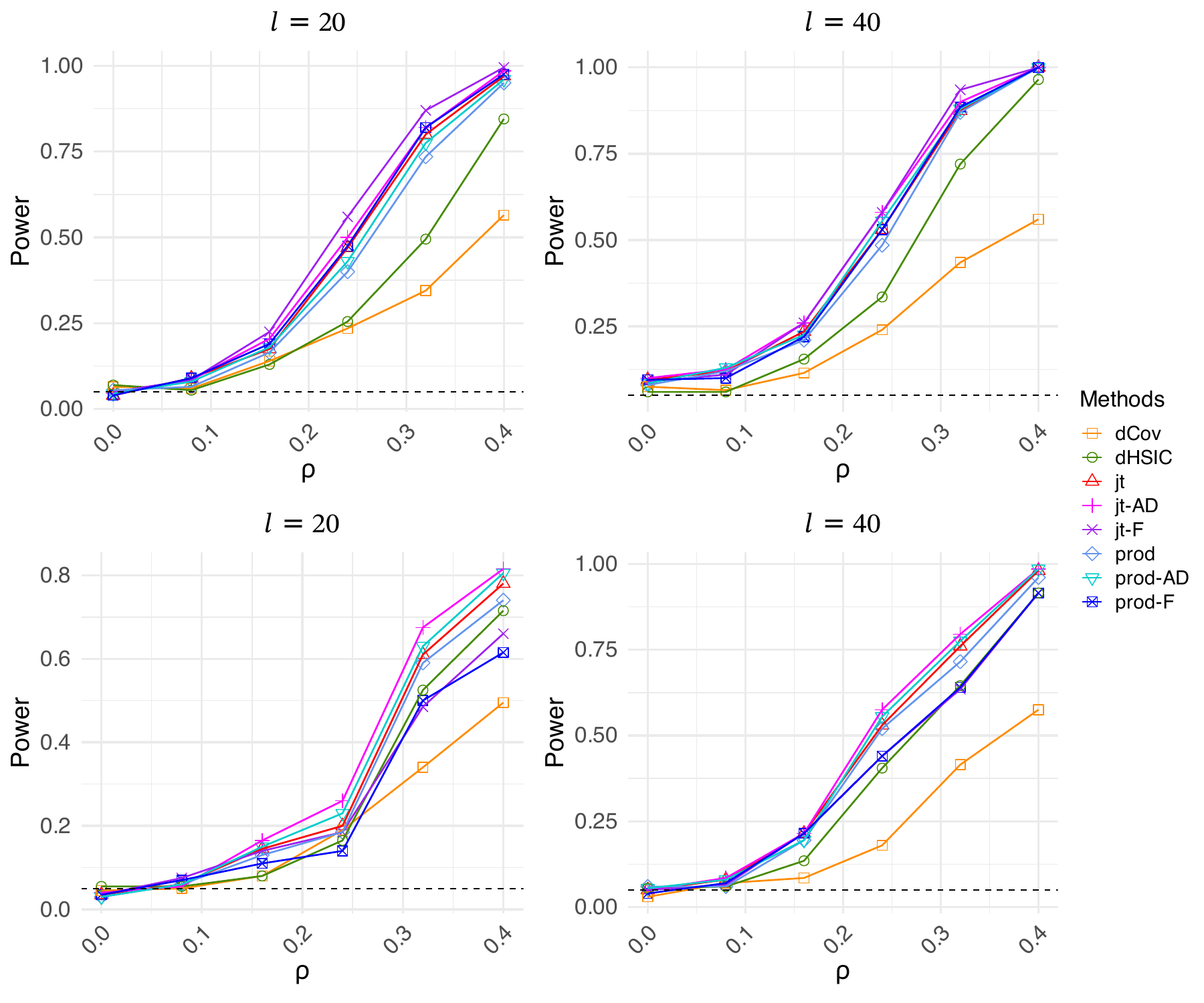}
 \caption{Empirical power as a function of $\rho$ for dimensions $l = 20$ and $l = 40$ for Example~\ref{ex:heavytailed}. In the top panel, data is generated as $\vrobj_i = (\robj_{i,1}, \robj_{i,2})\tps \in \mathbb{R}^{0.1l} \times \mathbb{R}^{0.9l}$, where $\rvecbm_i \in \mathbb{R}^l$ is sampled such that $\rvecbm_i \sim \gc{\mathbf{\Sigma}(\rho)_{l}}{(F, \dots, F)}$, with $F$ being the cdf of the standard Cauchy distribution. The components are defined as $\robj_{i,1} = (\rvec_{i,1}, \dots, \rvec_{i,0.1l})\tps$ and $\robj_{i,2} = (\rvec_{i,0.1l+1}, \dots, \rvec_{i,l})\tps$. In the bottom panel $F$ is the cdf of the $t$-distribution with two degrees of freedom. The components are defined as $\robj_{i,1} = (\rvec_{i,1}^2, \dots, \rvec_{i,0.1l}^2)\tps$ and $\robj_{i,2} = (\rvec_{i,0.1l+1}^2, \dots, \rvec_{i,l}^2)\tps$. The black dashed line indicates the significance level $\alpha = 0.05$.}
 \label{fig:heavytailed}
\end{figure}

\begin{example} \label{ex:heavytailed}
First, for $i =1, \dots, n$, we generate $\vrobj_{i} = (\robj_{i,1},\robj_{i,2})\tps \in \mathbb{R}^{0.1l} \times \mathbb{R}^{0.9l}$ by getting $\rvecbm_i \in \mathbb{R}^l$ such that $\rvecbm_i \sim \gc{\mathbf{\Sigma}(\rho)_{l}}{(F,\dots,F)}$, with $F$ being the cdf of the standard Cauchy distribution, i.e., $t$-distribution with one degree of freedom, and then we set $\robj_{i,1}=( \rvec_{i,1}, \dots,  \rvec_{i,0.1l})\tps$, and $\robj_{i,2}=( \rvec_{i,0.1l+1}, \dots,  \rvec_{i,l})\tps$. In the top panel of Figure~\ref{fig:heavytailed}, we present the empirical power curves of the different tests as a function of $\rho$ when $l=20$ and $l=40$. Next we generate $\rvecbm_i \in \mathbb{R}^l$ such that $\rvecbm_i \sim \gc{\mathbf{\Sigma}(\rho)_{l}}{(F,\dots,F)}$, with $F$ being the cdf of the $t$-distribution with two degrees of freedom and set $\robj_{i,1}=( \rvec_{i,1}^2, \dots,  \rvec_{i,0.1l}^2)\tps$, and $\robj_{i,2}=( \rvec_{i,0.1l+1}^2, \dots,  \rvec_{i,l}^2)\tps$ and illustrate the results in the bottom panel of Figure~\ref{fig:heavytailed}. The proposed tests outperform the competitors in this example with heavy-tailed data.
\end{example}

Next, we explore mutual independence among $p = 3$ random objects. As before, we begin with the non-Euclidean setting, followed by the Euclidean setting.

\begin{figure}[!hbt]
 \centering
\includegraphics[width=\textwidth]{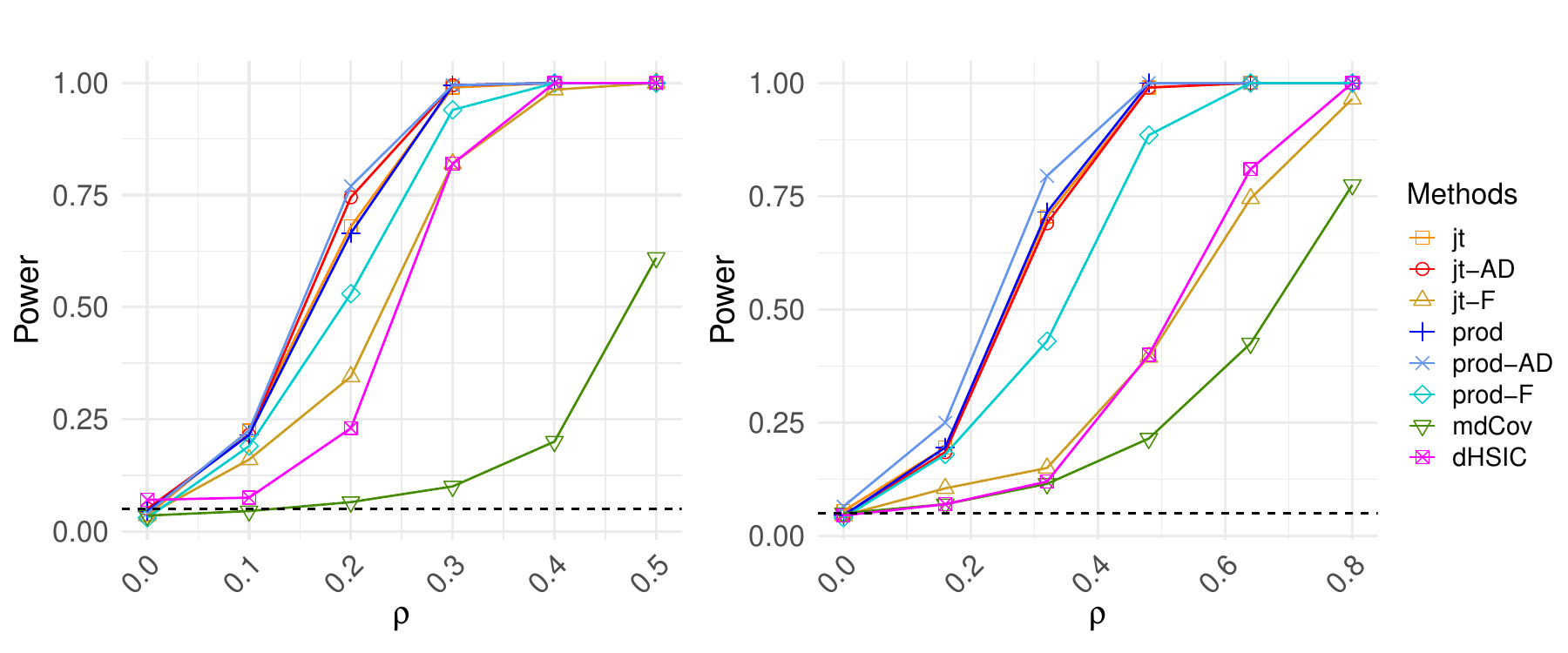}
 \caption{Empirical power as a function of $\rho$ for Setting 1 of Example~\ref{ex:mutual_NE} (left) and Setting 2 of Example~\ref{ex:mutual_NE} (right). The black dashed line indicates the significance level $\alpha = 0.05$.}
 \label{fig:mutual_NE}
\end{figure}

\begin{example}\label{ex:mutual_NE} We consider the following two settings.

\textit{Setting 1:} For $i =1, \dots, n$, we generate $\vrobj_{i} = (\robj_{i,1},\robj_{i,2},\robj_{i,3})\tps \in \msp_1 \times  \msp_2 \times \msp_3$ where $\msp_1 = \mathbb{S}^2=\{ \bm{x} \in \mathbb{R}^3 : \|\bm{x}\|=1\}$, the unit sphere in $\mathbb{R}^3$, and $\msp_2$ and $\msp_3$ are the spaces of graph Laplacian matrices of networks with $k_1=60$ and $k_2=100$ nodes, respectively. For $ \bm{x},\bm{y} \in \mathbb{S}^2$, we adopt the geodesic distance given by $d_{\mathbb{S}^2}(\bm{x},\bm{y})=\arccos(\bm{x}^\top \bm{y})$. We adopt the metric $d_k(\mathbf{L},\wt{\mathbf{L}}) = \sum_{j,j'=1}^{k} \lvert L_{j,j'}-\wt{L}_{j,j'} \rvert$ for $k\times k$ graph Laplacian matrices $\mathbf{L}$ and $\wt{\mathbf{L}}$. 
First we generate $\rvecbm_i \in \mathbb{R}^{15}$ such that $\rvecbm_i \sim \gc{\mathbf{\Sigma}(\rho)_{13,15}}{(F_1,\dots,F_{15})}$, with $F_j$ being the cdf of the $t$-distribution with two degrees of freedom for $j=1, \dots, 13$ and the cdf of $N(0,1)$ for $l=14,15$. We set $\robj_{i,1}=(V_{i,1}, V_{i,2}, V_{i,3})/(\sum_{j=1}^3 V_{i,j}^2)^{1/2}$. 
We generate the random networks $\robj_{i,2}$ (resp. $\robj_{i,3}$) according to the stochastic block model \citep{holl:83} with three blocks of sizes $30,15$ and $15$ nodes each (resp. five blocks each of size $20$ nodes). Here, the block connectivity matrices $\mathbf{A}_{i}$ and $\mathbf{B}_{i}$ for $\robj_{i,2}$ and $\robj_{i,3}$, respectively, are given by $\mathbf{A}_{i} = \diag\left(\frac{0.75}{1 + e^{-(\rvec_{i,4} \rvec_{i,5})}}, \frac{0.75}{1 + e^{-(\rvec_{i,6} \rvec_{i,7})}}, \frac{0.75}{1 + e^{- \{(2-\rvec_{i,8})^2 (2-\rvec_{i,9})^2\rvec_{i,10}\}}}\right) + 0.1 \mathbf{1}_{3}\mathbf{1}_{3}\tps$ and $\mathbf{B}_{i}=\diag\left(\frac{0.75}{1 + e^{-\rvec_{i,11}}},
    \frac{0.75}{1 + e^{-\rvec_{i,12}}},
\frac{0.75}{1 + e^{-\rvec_{i,13}}},
\frac{0.75}{1 + e^{-\rvec_{i,14}}},
\frac{0.75}{1 + e^{-\rvec_{i,15}}}
\right) + 0.05 \mathbf{1}_{5}\mathbf{1}_{5}\tps$. 
We present the empirical power curves of the different tests as a function of $\rho$ in the left panel of Figure~\ref{fig:mutual_NE}.

\textit{Setting 2:} For $i =1, \dots, n$, we generate $\vrobj_{i} = (\robj_{i,1},\robj_{i,2},\robj_{i,3})\tps \in \msp_1 \times  \msp_2 \times \mathbb{R}^{20}$ where $\msp_1$ is the space of compositional vectors in $\mathbb{R}^{20}$, that is, random vectors with non-negative entries that sum to one, and $\msp_2$ is the space of bivariate normal distributions, endowed with the 2-Wasserstein metric given by $W_2(N(\bm\mu_1, \mathbf{\Sigma}_1),N(\bm\mu_2, \mathbf{\Sigma}_2))=
\{\| \bm\mu_1 - \bm\mu_2 \|^2 + \| \mathbf{\Sigma}_1^{1/2} - \mathbf{\Sigma}_2^{1/2} \|_F^2\}^{1/2}
$ with $\| \mathbf{A} \|_F$ being the Frobenius norm of any matrix $\mathbf{A}$. We adopt the same metric for $\msp_1$ as described in Example~\ref{ex:compositionaldata}.
First we generate $\rvecbm_i \in \mathbb{R}^{60}$ such that $\rvecbm_i \sim \gc{\mathbf{\Sigma}(\rho)_{60}}{(F,\dots,F)}$, with $F$ being the cdf of the $t$-distribution with two degrees of freedom. We set $\robj_{i,1}=\operatorname{softmax}( \rvec_{i,1}^2, \dots, \rvec_{i,20}^2 )$ and $\robj_{i,3}=( \rvec_{i,41}, \dots,  \rvec_{i,60})\tps$. 
Next we generate $\robj_{i,2}$ to be a random bivariate normal distribution $N(\bm\mu_i, \mathbf{\Sigma}_i)$ such that $\bm\mu_i=( \sum_{j=21}^{25} \rvec_{i,j}, \sum_{j=26}^{30} \rvec_{i,j}
)\tps$ and $\mathbf{\Sigma}_i=\diag(
(1 + \exp\{-(\sum_{j=31}^{35}\rvec_{i,j})^2\})\inv,
(1 + \exp\{-(\sum_{j=36}^{40} \rvec_{i,j})^2\})\inv )$. 
We present the empirical power curves of the different tests as a function of $\rho$ in the right panel of Figure~\ref{fig:mutual_NE}. 

For both settings, the proposed tests have outstanding performance compared to the competitors. In particular, the \productAD versions of the proposed tests demonstrate the best performance, outperforming all other tests.
\end{example}

\begin{figure}[!hbt]
 \centering
\includegraphics[width=\textwidth]{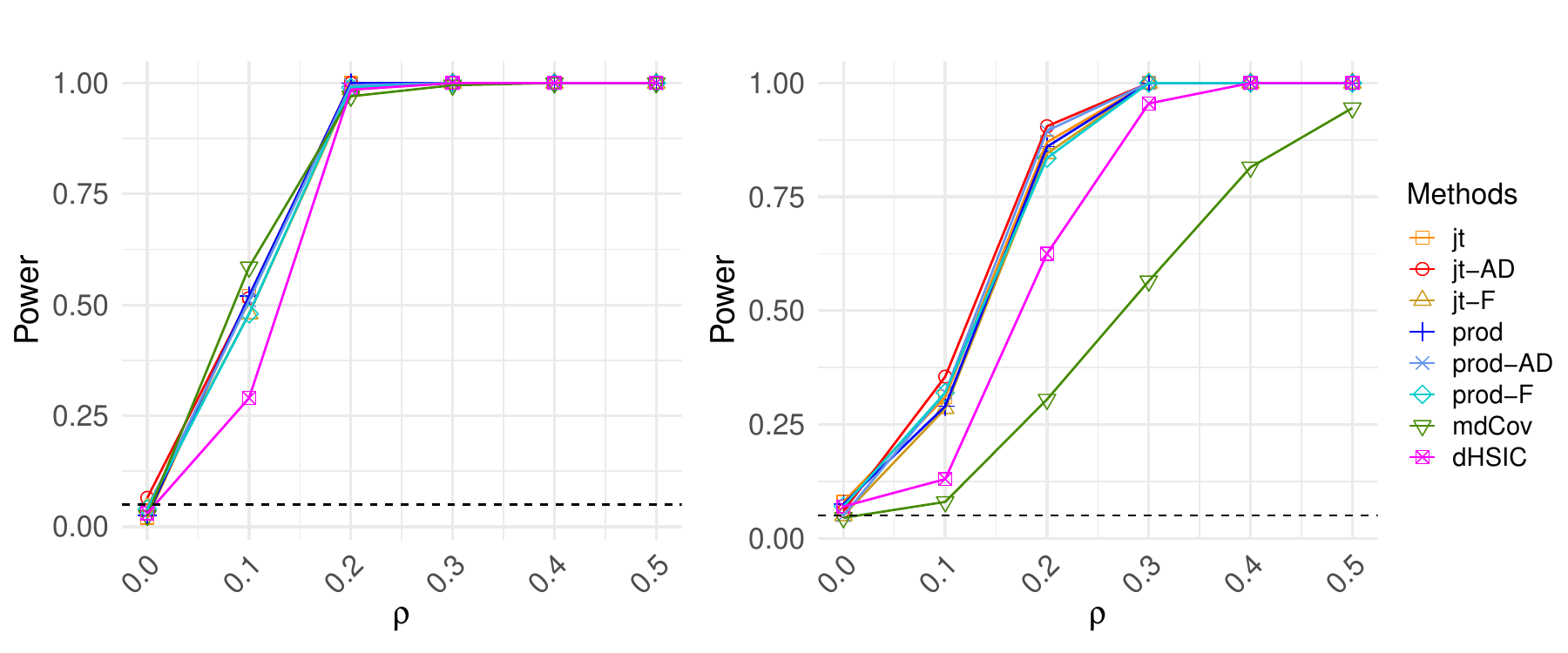}
 \caption{Empirical power as a function of $\rho$ for Setting 1 of Example~\ref{ex:mutual_euclidean_simple} (left) and Setting 2 of Example~\ref{ex:mutual_euclidean_simple} (right). The black dashed line indicates the significance level $\alpha = 0.05$.}
 \label{fig:mutual_E_simple}
\end{figure}

\begin{example}\label{ex:mutual_euclidean_simple}
For $i =1, \dots, n$, we generate $\vrobj_{i} = (\robj_{i,1},\robj_{i,2}, \robj_{i,3})\tps \in \mathbb{R}^{20} \times \mathbb{R}^{20} \times \mathbb{R}^{20}$ by getting $\rvecbm_i \in \mathbb{R}^{60}$ such that $\rvecbm_i \sim \gc{\mathbf{\Sigma}(\rho)_{60}}{(F,\dots,F)}$, with $F$ being the cdf of the $t$-distribution with degrees of freedom $\nu$ where $\nu=2$ in Setting 1 and $\nu=1$ in Setting 2, and then we set $\robj_{i,1}=( \rvec_{i,1}, \dots,  \rvec_{i,20})\tps$, $\robj_{i,2}=( \rvec_{i,21}, \dots,  \rvec_{i,40})\tps$ and $\robj_{i,3}=( \rvec_{i,41}, \dots,  \rvec_{i,60})\tps$. We present the empirical power curves of the different tests as a function of $\rho$ in Figure~\ref{fig:mutual_E_simple}. In Setting 1, the proposed tests perform comparably to \mdcov and outperform \dhsic, whereas in Setting 2, they outperform all competitors in the heavier-tailed data scenario.
\end{example}

\begin{figure}[!hbt]
 \centering
\includegraphics[width=\textwidth]{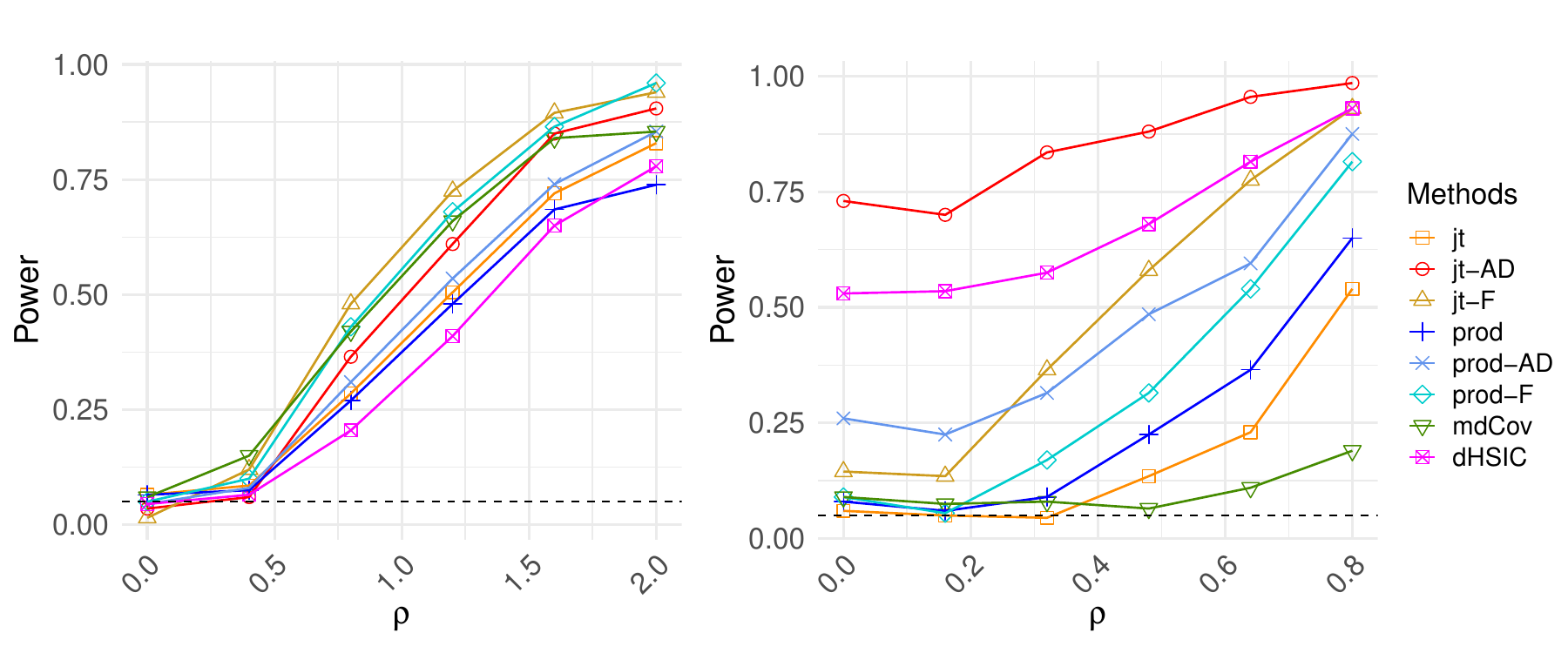}
 \caption{Empirical power as a function of $\rho$ for Setting 1 of Example~\ref{ex:mutual_euclidean_complex} (left) and Setting 2 of Example~\ref{ex:mutual_euclidean_complex} (right). The black dashed line indicates the significance level $\alpha = 0.05$.}
 \label{fig:mutual_E_complex}
\end{figure}

\begin{example}\label{ex:mutual_euclidean_complex} We consider the following two settings.

\textit{Setting 1:} For $i =1, \dots, n$, we generate $\vrobj_{i} = (\robj_{i,1},\robj_{i,2}, \robj_{i,3})\tps \in \mathbb{R}^{5} \times \mathbb{R}^{5} \times \mathbb{R}^{5}$ by getting $\rvecbm_i \in \mathbb{R}^{15}$ such that $\rvecbm_i \sim \gc{\mathbf{\Sigma}(0)_{15}}{(F,\dots,F)}$, with $F$ being the cdf of the $t$-distribution with two degrees of freedom, and then we set $\robj_{i,1}=( \rvec_{i,1}, \dots,  \rvec_{i,5})\tps$, $\robj_{i,2}=( \rho \log(\{1+e^{-{\rvec_{i,1}}}\}^{-1})+\varepsilon_{i,1}, \rvec_{i,6}, \dots,  \rvec_{i,10})\tps$ and $\robj_{i,3}=( \rho \log(\{1+e^{-{\rvec_{i,2}}}\}^{-1})+\varepsilon_{i,2}, \rvec_{i,12}, \dots,  \rvec_{i,15})\tps$, where $\varepsilon_{i,1}$ and $\varepsilon_{i,2}$ are generated independently from $N(0,1)$ .We present the empirical power curves of the different tests as a function of $\rho$ in the left panel of Figure~\ref{fig:mutual_E_complex}. The proposed tests, in particular the \jtF version, outperform all competitors in this scenario.

\textit{Setting 2:} For $i =1, \dots, n$, we generate $\vrobj_{i} = (\robj_{i,1},\robj_{i,2}, \robj_{i,3})\tps \in \mathbb{R}^{8} \times \mathbb{R}^{52} \times \mathbb{R}^{20}$. First we obtain $\rvecbm_i \in \mathbb{R}^{80}$ such that $\rvecbm_i \sim N(\mathbf{0}_{80},\mathbf{\Sigma}(\rho)_{16,64})$, and then we set $\robj_{i,1}=( \rvec_{i,1}, \dots,  \rvec_{i,8})\tps$, $\robj_{i,2}= ( \rvec_{i,9}, \dots,  \rvec_{i,60})\tps \mathbb{I}(|\rvec_{i,1}|+|\rvec_{i,2}| > 3) + \bm\varepsilon_{i,1} \mathbb{I}(|\rvec_{i,1}|+|\rvec_{i,2}| \leq 3)$ and $\robj_{i,3}=( \rvec_{i,61}, \dots,  \rvec_{i,80})\tps \mathbb{I}(|\rvec_{i,1}|+|\rvec_{i,2}| > 3) + \bm\varepsilon_{i,2} \mathbb{I}(|\rvec_{i,1}|+|\rvec_{i,2}| \leq 3)$, where $\bm \varepsilon_{i,1} \in \mathbb{R}^{52}$ and $\varepsilon_{i,2} \in \mathbb{R}^{20}$ are generated independently with each of their coordinates distributed independently as the $t$-distribution with two degrees of freedom. 
We present the empirical power curves of the different tests as a function of $\rho$ in the right panel of Figure~\ref{fig:mutual_E_complex}. 

In this example, the proposed tests and \dhsic outperform \mdcov, with the \jtAD variant of the proposed tests in particular dominating all other methods.
\end{example}

\begin{remark}
\label{rmk:simulation_assumptions_measure}
We demonstrate in the simulations that the proposed tests are powerful across many diverse scenarios in detecting mutual independence among random objects. 
All the examples investigated in the simulations satisfy either the condition in Eq.~\eqref{eq:limit_doubling} or the negative type condition in Proposition~\ref{prop:indep_measure}, and therefore the proposed measures \(\Ewpstat\) and \(\Ewjstat\) characterize mutual independence and the associated tests are valid at the population level.
\end{remark}

\begin{remark}
\label{rmk:simulation_assumptions_test}
For the proposed tests to achieve any desirable level under \(H_0\), we require Assumptions~\ref{ass:fcl_sjprf_brcktn_prime} and~\ref{ass:weight}. Remark~\ref{rmk:brcktn} provides broad, verifiable sufficient conditions for Assumptions~\ref{ass:fcl_sjprf_brcktn}/\ref{ass:fcl_sjprf_brcktn_prime}, illustrating that the bracketing condition requirement is flexible and accommodates a wide range of metric spaces and distributions. 
While Assumption~\ref{ass:fcl_sjprf_brcktn_prime} may be difficult to verify for some of the simulation scenarios, the proposed tests control Type~I error well across all the settings under $H_0$ and exhibit strong power performance throughout, indicating their robustness to possible departures from Assumption~\ref{ass:fcl_sjprf_brcktn_prime}. 
Regarding Assumption~\ref{ass:weight} on weight profiles, consider three weight profiles defined as: 
$\w_{\vobj}(\vrad) \equiv 1$, 
$\w_{\vobj}(\vrad) = \prod_{k=1}^{\nsp} (\max\{\eps, \mprf{k}_{\obj_{k}}(\rad_{k}) [1-\mprf{k}_{\obj_{k}}(\rad_{k})]\} )^{-1}$, 
and $\w_{\vobj}(\vrad) = \prod_{k=1}^{\nsp} (\max\{\eps, \mprf{k}_{\obj_{k}}(\rad_{k})\} )^{-1}$, 
for a constant $\eps>0$ which can be set arbitrarily small for the latter two. 
The corresponding estimated weight profiles for each $\vrobj_{i}$ are given by 
$\sw_{\vrobj_{i}}(\vrad) \equiv 1$, 
$\sw_{\vrobj_{i}}(\vrad) = \prod_{k=1}^{\nsp} (\max\{\eps, \smprf{k}_{\robj_{i,k}}(\rad_{k}) [1-\smprf{k}_{\robj_{i,k}}(\rad_{k})]\} )^{-1}$, 
and $\sw_{\vrobj_{i}}(\vrad) = \prod_{k=1}^{\nsp} (\max\{\eps,\smprf{k}_{\robj_{i,k}}(\rad_{k}) \})^{-1}$, respectively. 
All of these three satisfy Assumption~\ref{ass:weight}. 
In practice, taking $\eps \in (0,\frac{1}{n}(1-\frac{1}{n}))$, 
$\sw_{\vrobj_{i}}(\vrad) = \prod_{k=1}^{\nsp} (\max\{\eps, \smprf{k}_{\robj_{i,k}}(\rad_{k}) [1-\smprf{k}_{\robj_{i,k}}(\rad_{k})]\} )^{-1}$ and $\sw_{\vrobj_{i}}(\vrad) = \prod_{k=1}^{\nsp} (\max\{\eps, \smprf{k}_{\robj_{i,k}}(\rad_{k}) \} )^{-1}$ yield the same test statistics as $\sw_{\vrobj_{i}}(\vrad) = \prod_{k=1}^{\nsp} \{\smprf{k}_{\robj_{i,k}}(\rad_{k}) [1-\smprf{k}_{\robj_{i,k}}(\rad_{k})]\}^{-1}$ and $\sw_{\vrobj_{i}}(\vrad) = \prod_{k=1}^{\nsp} \{\smprf{k}_{\robj_{i,k}}(\rad_{k}) \}^{-1}$, respectively. 
Thus, in the beginning of Section~\ref{sec:simulations}, we simply define the empirical weight profiles using the latter simpler definitions without $\eps$. 
\end{remark}

\section{Data application}\label{sec:data}

In this data example, we study if hourly bike rental compositional vectors during a typical day are independent of daily temperature (\degC) distributions, humidity (\%) distributions, and wind speed (m/s) distributions across the seasons, while controlling for multiple comparisons. 
The dataset, publicly available at \url{https://archive.ics.uci.edu/dataset/560/seoul+bike+sharing+demand}, provides counts of public bicycles rented per hour in the Seoul Bike Sharing System together with weather data such as temperature (\degC), humidity (\%) and wind speed (m/s) recorded every hour from December 1, 2017 to November 30, 2018. 
We filter out the dates that correspond to holidays or include missing data. We normalize the hourly bike rental counts with the total number of bikes rented within each day and consider the resulting hourly bike rental compositions, for which we use the metric $d_c(\bm{y}_1,\bm{y}_2)=\arccos{(\sqrt{\bm{y}_1}\tps\sqrt{\bm{y}_2})}$ for any two compositional vectors $\bm{y}_1$ and $\bm{y}_2$ in $\real^{24}$, where $\sqrt{\bm{y}_1}$ and $\sqrt{\bm{y}_2}$ are entry-wise square root transformations. 
For temperature, humidity and wind speed, 
we use the hourly data per day to estimate the daily distributions, for which we adopt the $2$-Wasserstein metric given by the $L_2$-metric between the corresponding quantile functions. 
We conduct mutual independence tests between the hourly bike rental compositions and temperature distribution, humidity distribution, and wind speed distribution within each season and present our findings in Table~\ref{tab:season_distributions_highlighted}, where we highlight the permutation $p$-values in Eq.~\eqref{eq:ppval} obtained with $B=500$ that are significant at level $\alpha=0.05$ after Bonferroni corrections for the total number of tests across the seasons. We see that all versions of the proposed tests are more powerful in detecting relationships between the different random objects as compared to \dcov and \dhsic.

\begin{table}[!hbt]
\centering
\caption{$p$-values for the different tests of independence between hourly bike compositions and temperature (\degC) distribution, humidity (\%) distribution, and wind speed (m/s) distribution across the seasons. Significance after Bonferroni
correction for controlling multiple comparisons at level $\alpha =0.05/12$ is highlighted for each
test.}
\label{tab:season_distributions_highlighted}
\renewcommand{\arraystretch}{1.5}
\resizebox{\textwidth}{!}{%
\begin{tabular}{llllllll}
\toprule[1.25pt]
Season &                     Object &           \jt &           \jtAD &        \product &         \productAD &            \dcov &           \dhsic \\
\midrule
\midrule
Spring &  Temperature (\degC) Distribution & \textbf{0.0020} & \textbf{0.0020} & \textbf{0.0040} & \textbf{0.0020} &          0.0060 &          0.0240 \\
       &      Humidity (\%) Distribution & \textbf{0.0020} & \textbf{0.0020} & \textbf{0.0020} & \textbf{0.0020} & \textbf{0.0020} & \textbf{0.0020} \\
       & Wind speed (m/s) Distribution & \textbf{0.0040} & \textbf{0.0040} & \textbf{0.0040} & \textbf{0.0040} &          0.0419 &          0.1257 \\
\midrule
Summer &  Temperature (\degC) Distribution & \textbf{0.0020} & \textbf{0.0020} & \textbf{0.0020} & \textbf{0.0020} & \textbf{0.0020} & \textbf{0.0020} \\
       &      Humidity (\%) Distribution & \textbf{0.0020} & \textbf{0.0020} & \textbf{0.0020} & \textbf{0.0020} & \textbf{0.0020} & \textbf{0.0020} \\
       & Wind speed (m/s) Distribution & \textbf{0.0020} & \textbf{0.0020} & \textbf{0.0020} & \textbf{0.0020} & \textbf{0.0040} & \textbf{0.0020} \\
\midrule
Autumn &  Temperature (\degC) Distribution & \textbf{0.0020} & \textbf{0.0020} & \textbf{0.0020} & \textbf{0.0020} & \textbf{0.0020} &          0.0180 \\
       &      Humidity (\%) Distribution & \textbf{0.0020} & \textbf{0.0020} & \textbf{0.0020} & \textbf{0.0020} & \textbf{0.0020} & \textbf{0.0020} \\
       & Wind speed (m/s) Distribution &          0.0220 &          0.0240 &          0.0220 &          0.0220 &          0.0938 &          0.0758 \\
\midrule
Winter &  Temperature (\degC) Distribution &          0.0060 & \textbf{0.0040} &          0.0080 &          0.0060 &          0.0100 &          0.0599 \\
       &      Humidity (\%) Distribution & \textbf{0.0020} & \textbf{0.0020} & \textbf{0.0040} & \textbf{0.0020} &          0.0160 & \textbf{0.0040} \\
       & Wind speed (m/s) Distribution & \textbf{0.0040} & \textbf{0.0040} & \textbf{0.0040} & \textbf{0.0040} &          0.0399 &          0.1377 \\
\bottomrule[1.25pt]
\end{tabular}%
}
\end{table}

\section{Discussion}\label{sec:discussion}

In this paper, we propose a novel omnibus framework for testing the mutual independence of a vector of random objects with a joint law, which resides in the product of possibly different metric spaces. 
Our approach is founded on measuring the difference of joint distance profiles with respect to the joint law and the product of marginal laws, which equals zero under the null of mutual independence.
The proposed measures $\Ewpstat$ and $\Ewjstat$, as defined in Eq.~\eqref{eq:Ewjstat}, are measures of mutual independence under a mild doubling condition in Eq.~\eqref{eq:limit_doubling} on the law of random objects. 
Moreover, $\Ewpstat$ also characterizes mutual independence under other assumptions on the geometry of the underlying metric spaces, such as the assumption of Proposition~1 in \cite{dube:24}.

The proposed tests serve as alternatives to popular methods, such as distance covariance and dHSIC for metric spaces, offering comparable flexibility, such as the ability to test independence using only pairwise distances between data points but with superior empirical performance in various simulation instances, particularly in heavy-tailed settings and when the data distributions are corrupted with outliers. Specific choices of weight profiles recover well-known methods, including the ball covariance test \citep{pan:19:2,wang:24} and the test proposed in \cite{hell:13} as special cases; moreover, this paper provides a rigorous theoretical framework for the broader class of tests in separable metric spaces, along with new versions that exhibit better empirical performance across many scenarios. 
We establish the asymptotic null distribution of the proposed tests using tools from empirical process theory and $U$-process theory and ensure consistency of the test using approximated permutation $p$-values under the null. Moreover, we obtain power guarantees for both the asymptotic and the permutation test under sequences of alternatives converging to the null and establish that the proposed tests achieve minimax rate optimality.  
For trivial weight profiles, we show that the proposed tests are asymptotically distribution-free, extending the arsenal of distribution-free independence tests from random vectors to random objects. To the best of our knowledge, the proposed tests are the first asymptotically distribution-free tests for mutual independence for general metric-space-valued random objects.

For the proposed framework, potential follow-up research includes designing independence tests by measuring the difference of joint distance profiles and their values under the null using representations, e.g., characteristic functions or quantile functions, other than the cdf representation in Eq.~\eqref{eq:joint_prof}. 
Other future directions include investigating properties of the proposed tests for mutual independence among high-dimensional Euclidean data, as well as the robustness of the tests against outliers, a phenomenon validated empirically in Example~\ref{ex:outlier}.

%
%

\begin{acks}[Acknowledgments]
The authors would like to thank two anonymous referees, an associate editor, and the editor for their helpful and constructive comments which led to numerous improvements in the article.
\end{acks}
\begin{funding}
This research was supported in part by NSF grants DMS-2311035 (YC) and DMS-2311034 (PD).
\end{funding}

\begin{supplement}
The supplement contains proofs and auxiliary results for the asymptotic theory of the proposed tests and additional simulations.
\end{supplement}




\references
\end{document}